\documentclass[twocolumn,showpacs,amsmath,amssymb]{revtex4}

\usepackage{graphicx}
\usepackage{dcolumn}
\usepackage{bm}
\usepackage{amstext}

\usepackage{mymacros}

\begin{document}

\bibliographystyle{apsrev}

\preprint{Submitted to Physical Review B}

\title{
Longitudinal relaxation and thermoactivation of quantum
superparamagnets
}

\author{D. Zueco and J. L. Garc\'{\i}a-Palacios}

\affiliation{Departamento de F\'{\i}sica de la Materia Condensada e
  Instituto de Ciencia de Materiales de Arag\'on
\\
C.S.I.C -Universidad de Zaragoza, E-50009 Zaragoza, Spain}

\date{\today}

\begin{abstract}
The relaxation mechanisms of a quantum nanomagnet are discussed in the
frame of linear response theory.
We use a spin Hamiltonian with a uniaxial potential barrier plus a
Zeeman term.
The spin, having arbitrary $S$, is coupled to a bosonic environment.
From the eigenstructure of the relaxation matrix, we identify two main
mechanisms, namely, thermal activation over the barrier, with a time
scale $\eival_1^{-1}$, and a faster dynamics inside the potential
wells, with characteristic time $\eivalW^{-1}$.
This allows to introduce a simple analytical formula for the response,
which agrees well with the exact numerical results, and cover
experiments even under moderate to strong fields in the
superparamagnetic range.
In passing, we generalize known classical results for a number of
quantities (e.g., integral relaxation times, initial decay time,
Kramers rate), results that are recovered in the limit $S\to\infty$.
\end{abstract}

\pacs{03.65.Yz, 05.40.-a, 75.50.Xx, 75.50.Tt}


\maketitle 


\section{INTRODUCTION}
\label{sec:Introduction}

Superparamagnets are nanoscale magnetic solids or clusters effectively
described in terms of their net spin.
They have an internal anisotropy barrier and several local equilibrium
states.
The spin is coupled to the environmental degrees of freedom of the
host material (phonons, nuclear spins, electrons, \dots).
This coupling provokes disturbances in the spin dynamics producing
jumps over the barrier, thus magnetization reversals, giving rise to
the phenomenon of {\it superparamagnetism}.
The prefix {\it super\/} stands for the typically large net spins of
these systems ($S\sim10^1$--$10^4$).

Single-domain magnetic particles are an example of such nanoscale
solids, having a magnetic moment of a few thousand Bohr magnetons
\cite{panpol93}.
Due to their enormous spin these particles can be described
classically \cite{nee49,bro63}.
%
%
Another example is provided by single-molecule magnets, such as
Mn$_{12}$, Fe$_{8}$, or Ni$_{12}$ \cite{blupra04}.
%
%
Their net spin is $S\sim10$, so that the classical description is no
longer valid, and a quantum treatment is required.
%
%
Still, these molecules comprise some hundreds of atoms, yielding an
interplay of quantumness and mesoscopicity that has stimulated an
active research in recent years.
%

Classically, the spin surmounts the barrier $\Delta U$ by thermal
activation.
When $\Delta U/ \kT \gg 1$ the characteristic time for the overbarrier
process can be approximately written in the Arrhenius form
$\tau \equiv \eival_1^{-1} \propto \exp (\Delta U/ \kT)$.
This is the rotational countenpart of the Kramers' rate in the theory
of activated processes in translational systems
\cite{hantalbor90,mel91}.
The classical dynamics of these systems has been studied extensively.
In particular, the analysis of {\em longitudinal\/} response revealed
that the one-mode relaxation picture with a characteristic time
$\eival_1^{-1}$ is not valid in general \cite{cofetal95,gar96}.
A satisfactory description, however, is provided by a two mode
response, one mode corresponding to the overbarrier flux and the
other, with a time scale $\eivalW^{-1}$, describing the faster
intrawell dynamics \cite{kalcoftit03}.

Quantum mechanically, the spin reversal can take place by thermal
activation or tunneling (or a combination of both)
\cite{harpolvil96,garchu97,luibarfer98,wur98}.
Tunneling occurs from one side of the barrier to the other
(Fig.~\ref{fig:energy-levels}) between resonant states ``coupled'' by
transverse fields (or high-order anisotropy terms).
However in the superparamagnetic regime ($T \geq 2$\,K) thermal
activation controls the physics: out-of-resonance, the crossings are
only driven by thermal activation, while tunneling in resonance has
the effect of lowering the barrier a few states.

The theory of quantum thermal activation has focused mainly on the
zero, or very small bias limit \cite{viletal94, wur98prl}.
Then the relaxation becomes well described by one mode, the
overbarrier process.
The effect of bias fields has been only studied on the
thermoactivation rate \cite{gar97}, but not on the susceptibility, beyond the
weak-field regime.
Thus studies of the full dynamical response including the effect of
external fields are demanded.
This is the issue we address in this work: the relaxation of a spin
with arbitrary $S$ in contact with a bosonic (phonon) environment.
The dynamics is studied in the frame of quantum master (balance)
equations.
We concentrate on the case of longitudinal applied fields, giving a
complete characterization of the relaxation mechanisms.
The discreteness of the levels plays an important role in some aspects of
the dynamics.
The connection with the classical theory is done taking the
$S\to\infty$ limit (whenever well defined) recovering known classical
results.

The paper is organized as follows.
In Sec.~\ref{sec:QME_A} we present the quantum balance equations.
The tools of linear-response theory employed and the eigenstructure of
the relaxation matrix are discussed in Sec.~\ref{sec:ALR}.
%
%
The relaxation eigenvalue spectrum suggest the introduction of a
two-mode description of the susceptibility (Sec.~\ref{sec:AA})
following closely the approach of Kalmykov and co-workers in the
classical limit \cite{kalcoftit03, kalcoftit04}.
%
%
The comparison between exact numerical results and such analytical
approximation is done in Sec.~\ref{sec:R}.
We also give a complete characterization of the response and
simplified expressions (both for the susceptibility and relaxation
times) in the range $\Delta U/\kT \sim 10$--$20$, experimentally the
most relevant for superparamagnets.
%
%
%
Technical details for some calculations are sent to the appendices.


\section {QUANTUM SPIN IN A BOSONIC BATH}
\label{sec:QME_A}

\subsection {Spin Hamiltonian}

\begin{figure}[!tbh]
\centerline{
\resizebox{9.cm}{!}{%
\includegraphics[angle=-90]{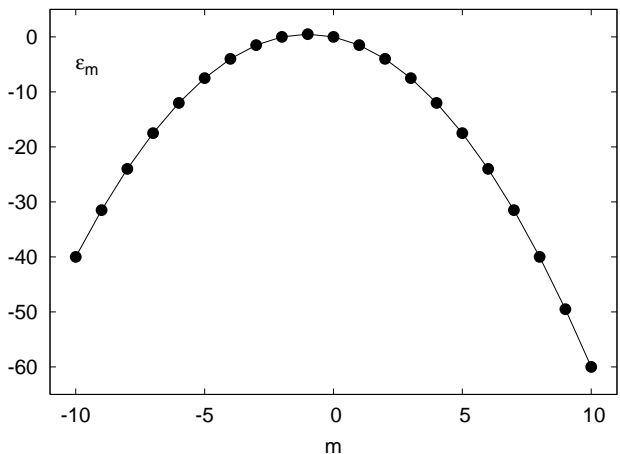}
}
}
\caption{
Energy levels for a spin $S=10$ described by the
Hamiltonian~(\ref{hamiltonian}) with $D=0.5$ at $\Bz=1$.
Considering that $D$ is given in Kelvin, the value used is closed to Mn$_{12}$.
}
\label{fig:energy-levels}
\end{figure}

The minimal Hamiltonian capturing the physics of superparamagnets
includes a uniaxial anisotropy term plus the Zeeman coupling with
external fields.
Here we study the longitudinal relaxation.
Thus only fields parallel to the anisotropy axis will be considered:
%
\begin{equation}
\label{hamiltonian}
\Hs
=
-D S_z^2 - B_z S_z
.
\end{equation}  
%
In the standard basis $\vl m \rangle$ this Hamiltonian provides a
spectrum $\varepsilon_{m}$ with a double well structure with minima at
$m=\pm S$ (for $B_z < D(2S-1)$; see Fig.~\ref{fig:energy-levels}).
%
%
The barrier heights are $\Delta U_{\pm} = \varepsilon_{\mb} -
\varepsilon_{\pm S}$, where $\mb$ is the index corresponding to the
maximum level \cite{fn:mb}
%
%
%
%
.
For fields $B_z \geq D(2S-1)$ the barrier disappears.
%

\subsection {Balance equations}

In contact with a bath, the dynamical equation for the diagonal terms
of the density matrix $\rho$ typically has a balance-equation form
\cite{cat90,gar97,garzue05}:
%
\begin{eqnarray}
\label{BE}
\nonumber
\dot N_m
=
\big (\Pmmplus N_{m+1}
-
\Pmplusm N_{m}
\big )
\\ 
+
\big (
\Pmmminus N_{m-1}
-
\Pmminusm N_{m}
\big )
,
\end{eqnarray}
%
where $N_m \equiv \rho_{mm} = \langle m \vl \rho \vl m \rangle$.
The transition probabilities $P_{ m \vl m'}$ depend on the energy
differences $\tf_ {m\,m'} \equiv \epsilon_m - \epsilon_{m'}$.
They fulfill the detailed balance condition:
%
\begin {equation}
\label {detailed}
\Pmmprima =
\e ^ {- \beta \tf_{m\,m'}} 
\Pmprimam
,
\end{equation}
which ensures that the Gibbs distribution, $N_m^{(0)} = \e^{- \beta
\Hs(m)}/ \mathcal Z$, would be a stationary solution of (\ref{BE}) ($\dot N_m = 0$).
The balance equations can be obtained in a phenomenological way:
postulating the form~(\ref{BE}) and then calculating the probabilities
$P_{m \vl m'}$ by means of Fermi's golden rule \cite{viletal94}.

Equation~(\ref{BE})  is a set of linear differential equations.
In matrix form it can be written as $\dot {\bf N} = - \R\,{\bf N}$,
with $[{\bf N}]_m \equiv N_m$ and $\R$ the relaxation matrix.
We assume $\R$ diagonalizable, and write the solution for $N_m$ as:
%
%
\begin{equation}
\label{generalsolutionBE}
{\bf N} = \sum_{i=1}^{2S}
c_i
\,
\e^{-\eival_i t}
\,
\vl \eival_i \rangle 
+
\vl \eival_0 \rangle 
\end{equation}
%
with $\eival_i$ the eigenvalues of $\R$ and $\vl \eival_i \rangle$ the
associated eigenvectors.
The last term in~(\ref{generalsolutionBE}) corresponds to the
equilibrium solution, i.e., $\vl \eival_0 \rangle_{m} \equiv
N_m^{(0)}$.
The relaxation matrix $\mathcal R$ must give ${\rm Re} (\eival_i) >
0$, ensuring that the asymptotic solution will be the equilibrium one
\cite{cat90}.

\subsection {Spin plus bath formulation}

Apart from phenomenologically, the balance equations~(\ref{BE}) can be
obtained in a more rigorous way \cite{cat90,gar97,garchu97}.
One starts from a total Hamiltonian representing the spin plus its
environment:
%
%
\begin{equation}
\label{totalHamiltonian}
\mathcal H_{ \rm tot}
=
\Hs(\s)
+
\sum_{\bi}
V_{\bi}
F_{\bi} (\s)
( a_{\bi}^+
+
a_{-{\bi}} )
+
\mathcal H_{\rm B}
.
\end{equation}
%
Here $\mathcal H_{\rm B} = \sum_{\bi} \omega_{\bi} a_{\bi}^+ a_{-\bi}$
is a bosonic bath modelling the host material, the $F_{\bi} (\s)$ the
spin-dependent part of the interaction, and $V_{\bi}$ coupling
constants.
%
%
%
If the spin-bath coupling is weak, the equation of motion for the
density matrix $\rho$ can be obtained within perturbation theory.
In absence of transverse fields, the diagonal elements of the equation
for $\rho$ form a closed set of kinetic equations like~(\ref{BE}),
with transition probabilities
%
\begin{equation}
\label{identification}
P_{m \vl m'}
= 
\vl 
L_{m, m'}
\vl ^2
W_{m \vl m'}
.
\end{equation}
The $L_{m, m'}$ are matrix elements of the spin-dependent part of the
coupling $F(\s)$ and the rates $W_{m \vl m'}\equiv W(\tf_{m\,m'})$,
are given in terms of the following universal function \cite{garzue05}
\begin{equation}
\label{W}
\Wu(\tf)
=
\frac {\lambda\,\tf^{\kk}} { \e^{ \beta \tf} -1}
\;.
\end{equation}
%
Here we have considered that the spectral density of the bath,
$J(\omega)=
\frac{\pi}{2}
\sum_{\bi}
|V_{\bi}|^{2}\,\delta(\omega-\omega_{\bi})$,
has the form $J(\omega) \propto \omega^{\kk}$ at low $\omega$.
When $\kk =1$ the bath is called {\it Ohmic}, while if $\kk >1$ it is
called {\it super-Ohmic}.

Having molecular magnets in mind, we will use the following coupling
\cite{harpolvil96,garchu97}:
$F(\s) \propto \big \{S_z,\,S_{\pm} \big \}$,
then
$ L_{m, m \pm 1} = (2m \pm 1) \sqrt {S(S+1) - m(m \pm 1)}$
\cite{garzue05}.
In addition the bath will be super-Ohmic ($ \kk = 3 $).
This models the interaction with 3D phonons considering one-phonon
emission and absorption process, the relevant ones at low temperatures
for single-molecule magnets.
The modifications required to include other structures of the
coupling, or other baths, would only involve the $P_{m \vl m'}$ and
they are easy to carry out.

\subsection{Classical limit}

As we would like to make the connection of our results with known
classical results, let us briefly consider the $S\to\infty$ limit of
the balance equations~(\ref{BE}).
To this end it is useful to introduce the scaled quantities:
%
\begin{equation}
\label{scaledxisigma}
\sigma
\equiv
\beta D S^2
\;,
\qquad
\xi
\equiv
\beta B_z S
\;,
\qquad
\hef 
\equiv
\frac {\xi}{\left(2 - \frac{1}{S} \right) \sigma}
\;.
\end{equation}
%
In terms of them the scaled Hamiltonian reads $\beta\Hs=-\sigma
(m/S)^2 - \xi (m/S)$.
The first two quantities are equivalent to those used in the
description of classical magnetic nanoparticles.
The ``reduced field'' $\hef$, is $B_z$ measured in terms of the field
for barrier disappearance $D(2S-1)$; it differs from the usual
classical definition $\hef_{{\rm cl}} = \xi/(2\sigma)$, as a
consequence of the discreteness of the energy levels.

%
The quantities $\sigma$ and $\xi$ should be kept constant when taking
the limit $S \to \infty$.
Then $\hef \to \hef_{{\rm cl}}$ and $\beta \mathcal H\to
-\sigma z^2 - \xi z \equiv u(z)$.
%
%
%
%
Physically more and more levels are introduced, towards a continuum,
while keeping the anisotropy barrier and Zeeman energy constant.
In this limit the transition frequencies $\tf_{m, m \pm 1}$ tend to
zero.
Thus, no relaxation is left in the classical case if $P_{m \vl m \pm
  1} (0) = 0$ (see App.~\ref{app:CL}).
Such is the case for a pure super-Ohmic environment.
A well defined classical limit ($P_{m \vl m \pm 1} (0)\neq 0$) is
obtained considering an {\it Ohmic} bath.
For instance, for coupling to electron-hole excitations or just adding
two phonon processes (Raman scattering) in the interaction to phonons.
%
%
%
In those cases, the balance equation (\ref{BE}) goes in the limit $S
\to \infty$ over a partial differential ({\em Fokker--Planck})
equation:
\begin{eqnarray}
\label{f-pclassical}
\frac {\partial \W}{\partial t}
=
\frac {\partial}{\partial z}
\Big [
D (z)
\Big (
\frac {\partial \W}{\partial z}
+
\frac {\partial u}{\partial z}
\W
\Big )
\Big ]
,
\end{eqnarray}
%
with $W (z,t)$ the probability distribution of $z$ ($\sim m/S$) and
$P_{m+1 \vl m}(0)/S^2 \to D(z)$ (see App.~\ref{app:CL} for the details).
%

%
%
%
%
%
%
%
%
 

\section{ANALYSIS OF THE LONGITUDINAL RESPONSE}
\label{sec:ALR}

The purpose of this section is to present the theoretical tools
necessary to describe the relaxation of a spin $S$.
To made a system to ``relax'' we must put it in a non-equilibrium
situation (e.g., subjecting it to a perturbation in the external
field).
The response to this probe informs about the relaxation mechanisms of
the system.
We will use linear-response-theory tools \cite{dattagupta} and the
analysis of the eigenstructure of the relaxation matrix.
%


\subsection {Linear response theory}

For the effect of the applied perturbation to reflect the intrinsic
properties of the system the force should be suitable small.
%
%
The weakness of the probe has the technical advantage of allowing the
use of linear response theory.
Several ``experiments'' can be made to study the response, namely,
subjecting it to a sudden constant ``force'' or by removing the force
after having kept it on for a long time.
One can also consider the response to a force oscillating with
frequency $\Omega$.
Linear response theory provides an intimate relationship between these
types of measurements.
%

Assume first that we have excited the system with a constant field
$\delta B_z$, keeping it on until the system is equilibrated (in a
total field $B_z^{0} + \delta B_z$).
Then, we switch the perturbation off and measure the response:
%
\begin{equation}
\Delta M_z (t)
\equiv
\langle M_z(t) \rangle
-
\langle M_z \rangle_0
,
\end{equation}
%
with $\langle M_z \rangle_0$ the statistical average at equilibrium
with $B_z = B_z^{0}$ and
%
%
$
\langle M_z (t) \rangle
=
\sum_{m=-S}^{S}
m
N_m(t)
$.
Since the asymptotic solution for ${\bf N}$ is the equilibrium one in
$B_z = B_z^{0}$ then $\langle M_z \rangle_0 = \langle M_z( t \to
\infty) \rangle$.
Next, introducing the solution~(\ref{generalsolutionBE}) for $N_m(t)$
in $\langle M_z(t) \rangle$, we find
%
\begin{equation}
\label{response1}
\Delta M_z
=
\suseq
\delta B_z
\sum_{i=1}^{2S}
\as_i
\e^{-\eival_i t}
\end{equation}
%
where $\suseq$ is the equilibrium susceptibility:
$\suseq\equiv \partial \langle M_z \rangle_0/\partial B_z$.
In linear response $\suseq$ is equal to
$[\langle M_z(0)\rangle-\langle M_z(\infty)\rangle]/\delta B_z$.
The coefficients $\as_i$ are given in terms of the eigenvalues of the
relaxation matrix $\R$ and the coefficients $c_i$ by [see
Eq.~(\ref{generalsolutionBE})]:
%
\begin{equation}
\label{as}
\as_i
=
\frac {1}{ \suseq}
\tilde{c}_i
\sum_{m=-S}^{S}
m
\vl \eival_i \rangle_m
.
\end{equation}
%
Here $c_i$ has been redefined as $\tilde{c}_i \equiv c_i/ \delta B_z$,
to get rid of dependences on $\delta B_z$.
The coefficients $\as_i$ obey the normalization condition:
$\sum_{i=1}^{2S}\as_i=1$ which follows from the definition (\ref{as})
and the equality: $\suseq = \sum_{i} \tilde {c}_i \sum_{m} m \vl \eival_i
\rangle_m $.
The $\tilde{c}_i$ can be evaluated making $t=0$ in
Eq.~(\ref{generalsolutionBE}).
The initial condition is the system equilibrated at $B = B_z^{0} +
\delta B_z$.
%
The occupation numbers can be written in linear approximation as
$N_m (t=0) \cong N_m^{(0)} B_z ^ {(0)} + \delta B_z \partial_{B_z}
N_m^{(0)} \vl_{B_z^{0}}$,
so that the $\tilde{c}_i$'s obey
%
\begin{equation}
\label{c_i}
\frac {\partial N_m^{{\rm eq}}} {\partial B_z} 
\bigg \vl_{B_z^{0}}
=
\sum_{i=1}^{2S}
\tilde{c}_i \vl \eival_i \rangle_m
, 
\end{equation}
%
which is an overdetermined set of $2S +1$ linear equations (there are
$2S$ $c_i$'s).
Both $\vl \eival_1 \rangle$ and $\tilde{c}_i$ can be obtained using
standard numerical routines \cite{lapack}.


Finally, let us consider the relation with the other relaxation
experiments.
The case with the system equilibrated at $B_z = B_z^{0}$ and a
perturbation $\delta B_z$ suddenly added is just the complementary
situation to case discussed above.
More interesting is the oscillating field probe.
Here we define the dynamical susceptibility, $\sus$, as the
coefficient which relates (in the frequency domain) response and
excitation:
%
$
\Delta \widetilde M_z (\Omega)
=
\sus
\widetilde {\delta B_z} (\Omega)
$
with $\widetilde g(\Omega) = \int \D t\, \e^{-\iu\Omega t}g(t)$.
Then the response to a periodic perturbation $\delta B_z (t) \propto
\e^{-\iu\Omega t}$ is given by
%
\begin{eqnarray}
\label{dynamicalsus}
\nonumber
\sus &=&
\suseq 
\Big [
1
-
\iu\Omega
\int_{0}^{\infty}
\!
\D t\,
\,
\frac {\Delta \Mz (t)}{\Delta \Mz (0)}
\e^{-\iu\Omega t}
\Big ]
\\
&=&
\suseq
\sum_{i=1}^{2S}
\frac {\as_i}
{1 + \iu\Omega \Lambda_i^{-1}}
\end{eqnarray}
%
Equations~(\ref{response1}) and~(\ref{dynamicalsus}) provide the
relation between the different relaxation experiments.
%
%
The analysis of the response gives the time scales $\eival_i^{-1}$ and
the weights $\as_i$'s of the modes involved in the relaxation process.
To conclude, note that neither $\eival_i$ nor $\as_i$ depend on the
external probe but only on the intrinsic dynamics of the system, as it
was demanded.


\subsection{ Eigenvalue and eigenvector structure}
\label {subsec:EES}

\begin{figure}[!tbh]
\centerline{
\resizebox{9.cm}{!}{%
\includegraphics[angle=-90]{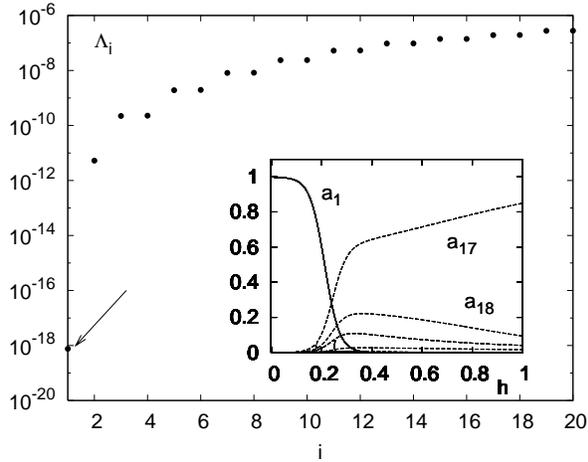}
}
}
%
%
\caption{
Eigenvalues of the relaxation matrix for $S=10$.
The spin-bath coupling is $\lambda=10^{-9}$ , $\sigma=\beta D
S^2=15$ and $h = 10 ^{-3}$.
The arrow marks the lowest non-vanishing eigenvalue $\eival_1$.
Inset: coefficients $\as_i$ as a function of $\hef$ for the same $S$,
$\lambda$, and $\sigma$.
}
\label{fig:eigenvalues}
\end{figure}

\begin{figure}[!tbh]
\centerline{
\resizebox{9.5cm}{!}{%
\includegraphics[angle=-90]{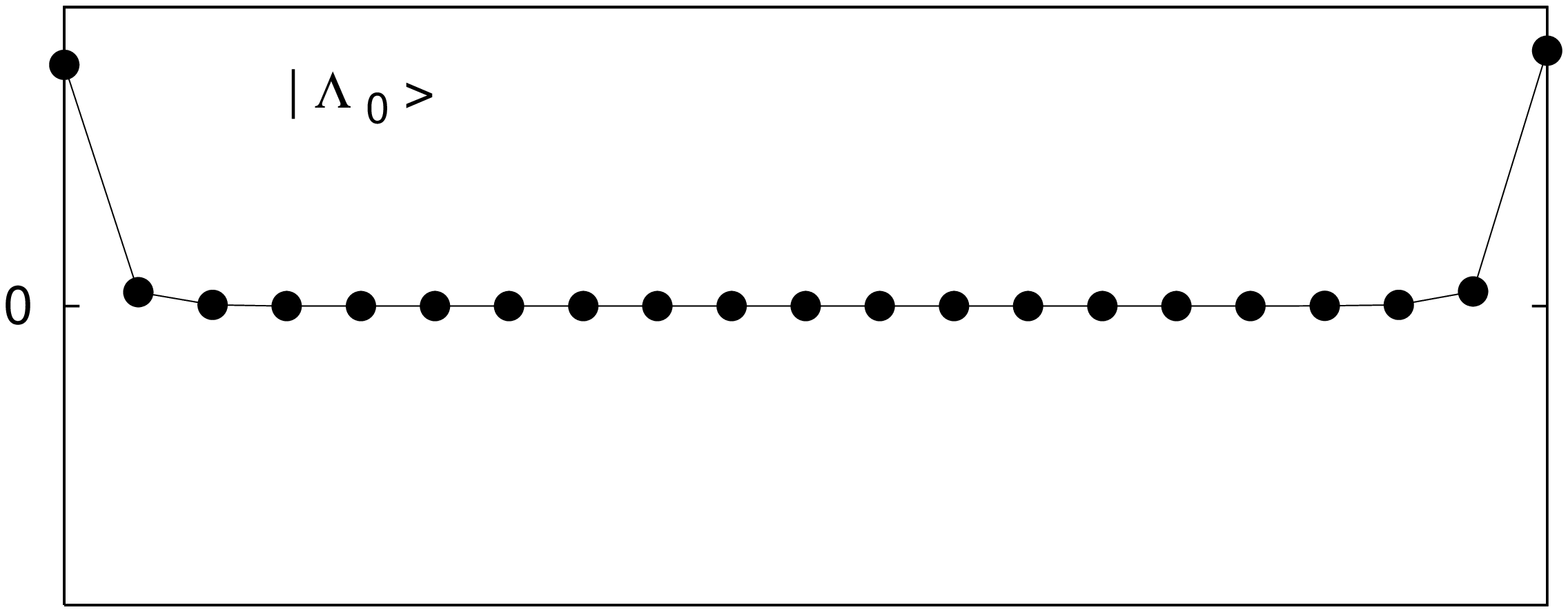}
}
}
\vspace*{-4.ex}
\centerline{
\resizebox{9.5cm}{!}{%
\includegraphics[angle=-90]{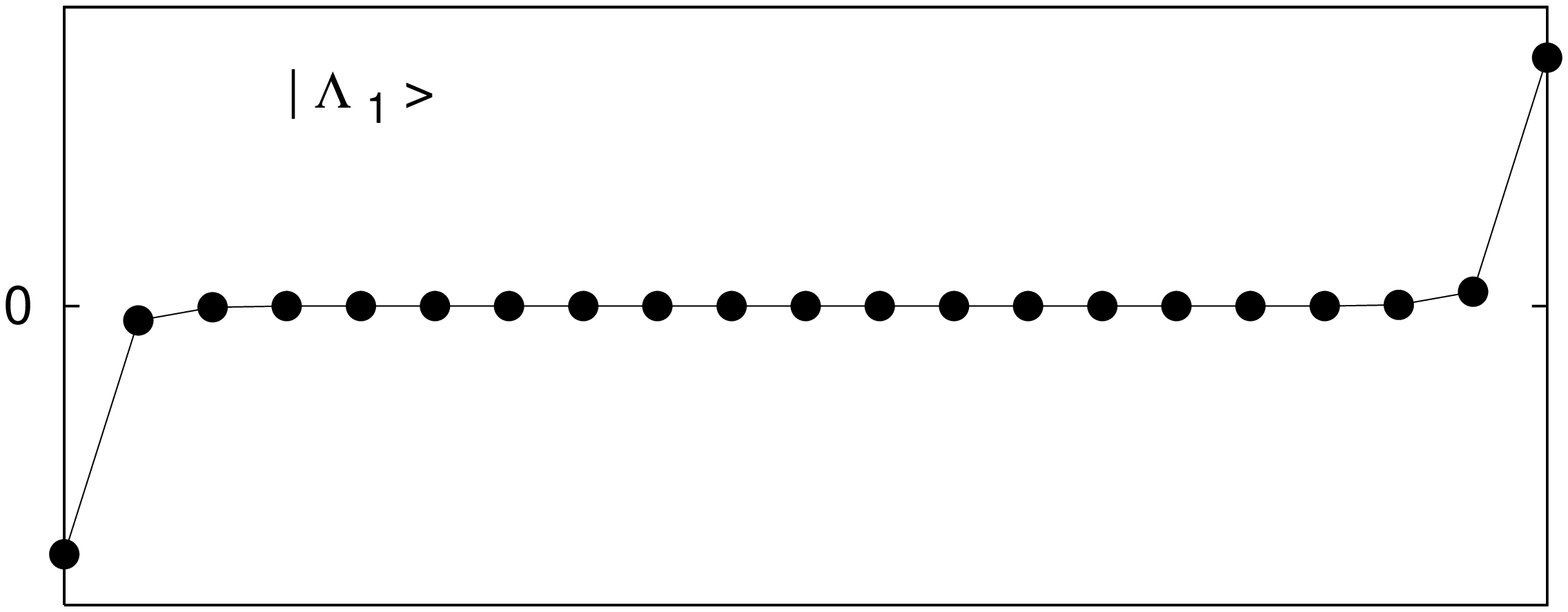}
}
}
\vspace*{-4.ex}
\centerline{
\resizebox{9.5cm}{!}{%
\includegraphics[angle=-90]{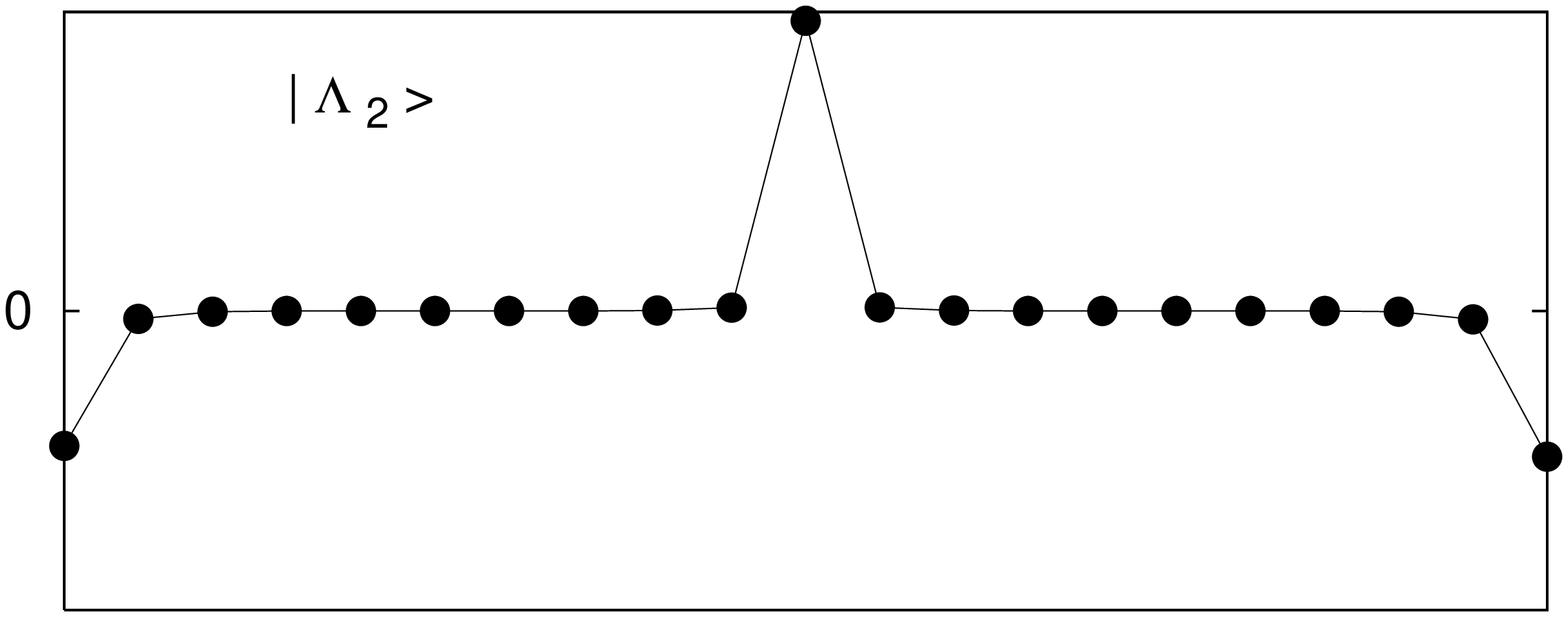}
}
}
\vspace*{-4.ex}
\centerline{
\resizebox{9.5cm}{!}{%
\includegraphics[angle=-90]{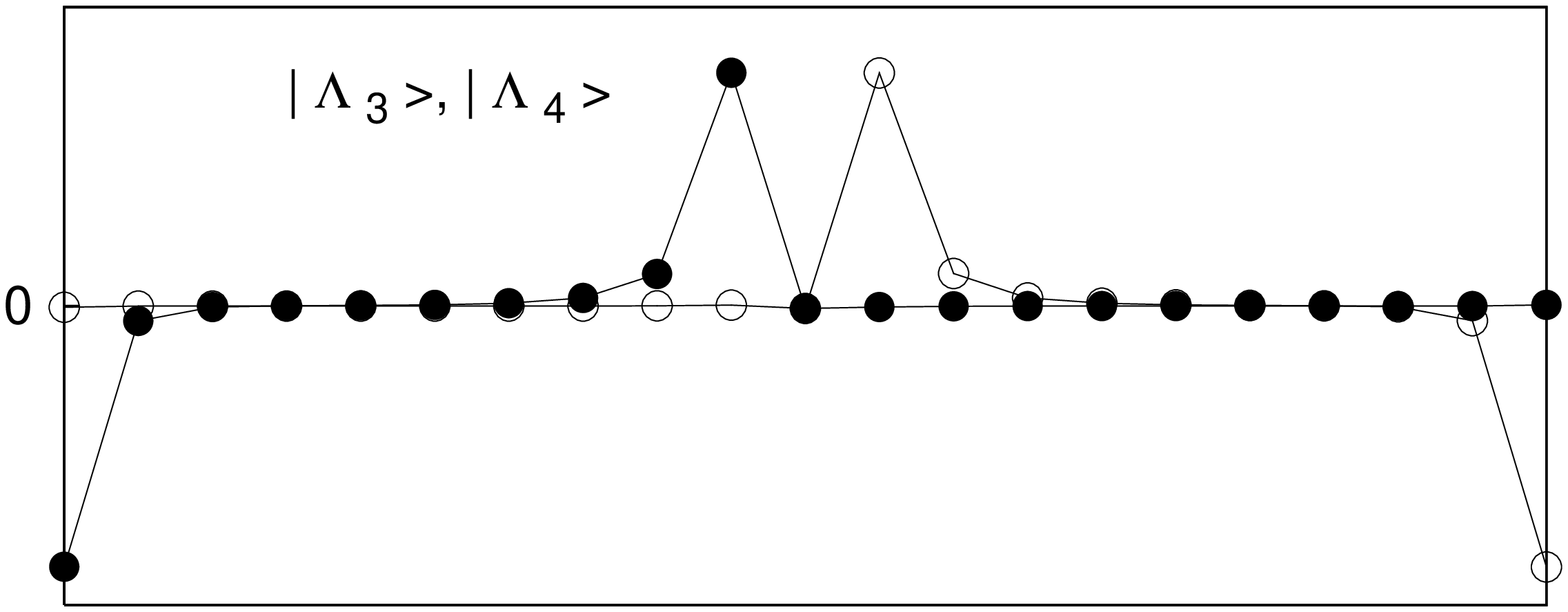}
}
}
\vspace*{-4.ex}
\centerline{
\resizebox{9.5cm}{!}{%
\includegraphics[angle=-90]{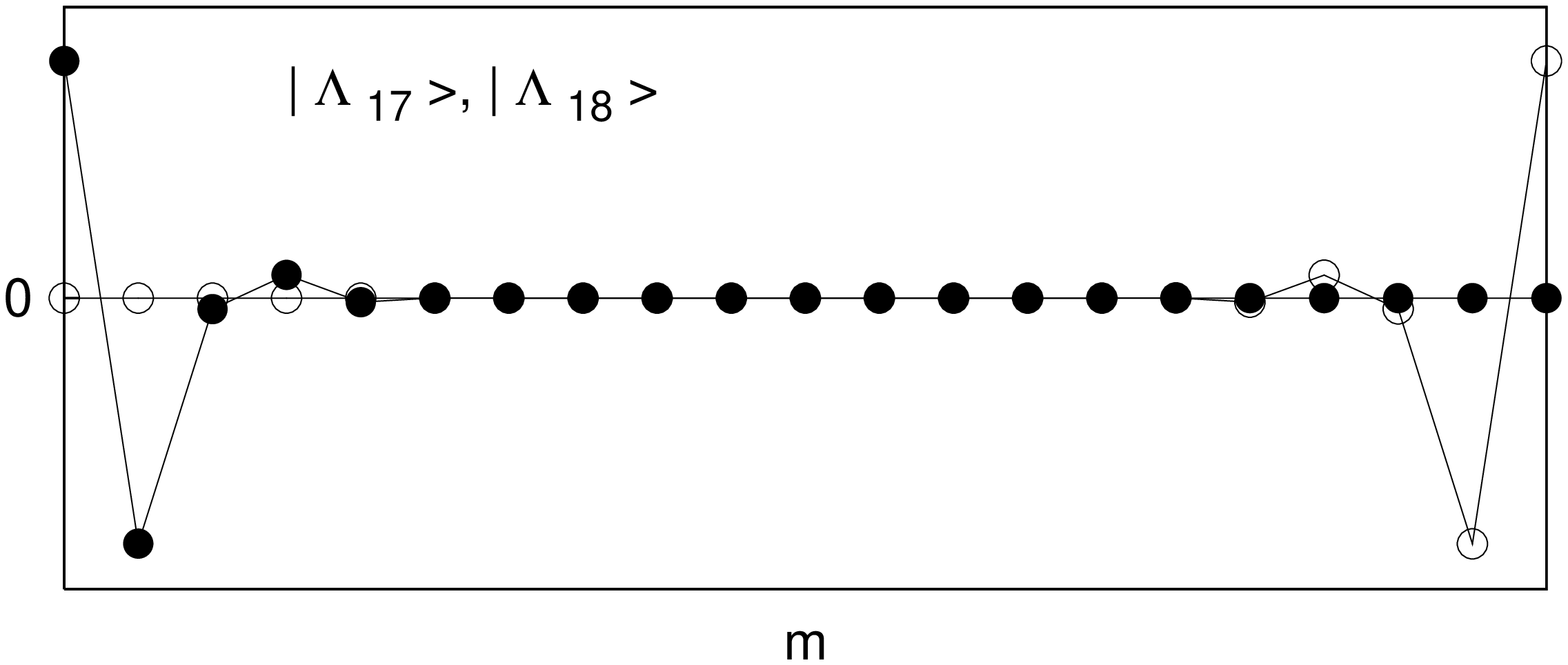}
}
}
\caption{
Eigenvectors of the relaxation matrix for $S=10$, $\sigma =15$,
$\hef=10^{-3}$ and $\lambda=10^{-9}$.
%
%
At $\hef = 0$ the states $\eival_{2m -1}$ and $\eival_{2m}$ ($m \geq 2$)
are degenerated.
Numerical diagonalization gives even-odd eigenvectors, so we use a
small $\hef=10^{-3}$.
%
}
\label{fig:eigenvectors}
\end{figure}

Attending at the evolution of ${\bf N}$ in
Eq.~(\ref{generalsolutionBE}), we see that the eigenvalues of $\R$
give the different time scales $\eival_i^{-1}$ in the relaxation,
while the eigenvectors provide the change in the population levels.
The eigenvalues and eigenvectors for $S=10$, obtained by numerical
diagonalization of $\R$, are plotted in Figs.~\ref{fig:eigenvalues}
and \ref{fig:eigenvectors}.
%
%

Apart from the zero eigenvalue (not plotted in
Fig.~\ref{fig:eigenvalues}), the eigenvalues correspond to two, well
separated, sets of time scales.
From the structure of the eigenvectors (Fig.~\ref{fig:eigenvectors})
we see that the dynamics induced by $\vl \eival_1 \rangle$ produces
the increment in the population in one well and the depopulation of
the other.
Since this transfer occurs across the barrier, $\vl \eival_1 \rangle $
is associated to the overbarrier dynamics.
On the other hand, $\vl \eival_2 \rangle$ represents the transfer
from $m=0$ to both wells or viceversa.
Further, $\vl \eival_3 \rangle$ and $\vl \eival_4 \rangle$ account for
dynamics involving $m = \pm 1$ and $m =\pm S$ respectively, and so on.
Eventually $\vl \eival_{17}\rangle$--$\vl \eival_{20=2S}\rangle$,
involve the levels close the bottom of the wells.
All these processes are called intrawell ones, since no activation
over the barrier is required.
Thus the slow, and well separated, time scale $\eival_1^{-1}$
corresponds to the overbarrier dynamics, whereas the set of
(comparatively close) intrawell processes is responsible for
the fast dynamics.

Let us briefly comment on two aspects of the eigenstructure.
First, apart from $\eival_0$, $\eival_1$ and $\eival_2$, the remaining
eigenvalues appear doubly degenerated (at $\hef=0$) \cite{fn:hi}
%
%
%
%
%
.
Physically each member of the degenerate pair describes identical
processes in different wells, and at $B_z=0$ both wells are
equivalent.
When $\Bz \neq 0$ the equivalence is broken and the degeneration
lifted.

Finally a connection with the classical case can be sketched.
The {\it Fokker--Planck\/} limit~(\ref{f-pclassical}) of the balance
equations~(\ref{BE}) defines a Sturm-Liouville problem.
From the corresponding theory we know that the eigenvalues of the
differential operator are real and we can order them:
$\eival_0^{{\rm cl}} \leq \eival_1^{{\rm cl}} \leq \eival_2^{{\rm
cl}}\leq\dots$.
Their corresponding eigenvectors $\vl \eival^{{\rm cl}}_n \rangle$
have $n$ nodes.
Restricting ourselves to  $\hef = 0$, the problem is invariant
under the transformation $z \to -z$.
Then there exists a basis where the states are even or odd under $z
\to -z$.
Then $\vl \eival_0^{{\rm cl}} \rangle$ (the Gibbs solution) is an
even function with zero nodes.
Next $\vl \eival_1^{{\rm cl}} \rangle $, would be an odd function
since it has $1$ node, $\vl \eival_2^{{\rm cl}} \rangle$ an even
function (with $2$ nodes), and so on.
These features are retained in the discrete ($S<\infty$) case
(Fig.~\ref{fig:eigenvectors}).
Actually, from the degenerate states (e.g., $\vl \eival_3 \rangle$ and
$\vl \eival_4 \rangle$) we can form the symmetrical and
antisymmetrical combinations $ \vl \eival_3 \rangle \pm \vl \eival_4
\rangle$, fully recovering the even-odd picture.


\section {ANALYTICAL APPROACH}
\label{sec:AA}

In general, the response will depend on $2 S$ modes [see Eqs. (\ref{response1})
and (\ref{dynamicalsus})].
However, taking into account
the eigenvalue structure of the previous section, 
we approximate the
response in terms of two main processes (see Sec.~\ref{subsec:EES}).
%
%
%
One accounting for the overbarrier dynamics, with characteristic time
$\eival_1^{-1}$, and the other describing the intrawell dynamics with
a time scale $\eivalW^{-1}$.
The latter is a kind of {\em collective\/} mode which describes the
close set of intrawell processes.
This motivation for the two-mode approximation is reinforced by the
good results it yields in the classical limit
\cite{kalcoftit03,kalcoftit04}.


\subsection {Bimodal approximation}

Suppose that the relaxation can be approximated by two exponentials:
%
\begin{equation}
\label{relaxationapprox}
\Delta \Mz (t)
\cong
\suseq \delta B_z
\big[
\Deltauno
\e^{-\eival_{1} t}
+
(1- \Deltauno)
\e^{-\eivalW t}
\big]
,
\end{equation}
or, in the frequency domain,
%
\begin{equation}
\label{susapprox}
\frac {\sus}{\suseq}
\cong
\frac {\Deltauno}{1 + \iu \Omega \eival_{1}^{-1}}
+
\frac {(1 - \Deltauno)}{1 + \iu \Omega \eivalW^{-1}}
.
\end{equation}
Note first that this bimodal description is exact for $S=1$
(see App.~\ref{app:S1}).
In general $\Deltauno$ and $\eivalW$ are coefficients to be identified
with known magnitudes of the problem.
This can be carried out following the approach of Kalmykov and
co-workers in the classical case \cite{kalcoftit03}.
They observed that we have two unknown parameters ($\Deltauno$ and
$\eivalW$), while $\sus$ can be evaluated in two extreme limits,
namely, the low-$\Omega$ and high-$\Omega$ ranges.
Then, taking these limits in the above two mode approximation,
$\Deltauno$ and $\eivalW$ are identified.

The low frequency behavior of $\sus$ is given by:
%
%
\begin{eqnarray}
\label{suslowomega}
\sus
\cong
\suseq
(
1
-
\iu\Omega
\tint
+
\cdots
)
\;,
\end{eqnarray}
with $\tint$ the integral relaxation time, defined as the area under
the magnetization relaxation curve after a sudden change $\delta B_z$
at $t=0$:
%
\begin{equation}
\label{tint}
\tint
\equiv
\int_{0}^{\infty}
\!
\D t\,
\,
\frac {\Delta \Mz (t)}{\Delta \Mz (0)}
\;.
\end{equation}
%
Next, in the high-$\Omega$ limit the susceptibility can be expanded
as:
%
\begin{equation}
\label{sushighomega}
\sus
\cong
\suseq
\frac {\iu}{\Omega \tef}
\;,
\qquad
\tef^{-1}
\equiv
\frac {\D}{\D t}
\left (
\frac {\Delta \Mz (t)}{\Delta \Mz (0)}
\right )
\bigg |_{t=0}
\;,
\end{equation}
%
where $\tef$ is the initial slope of the relaxation.

The proposed formula~(\ref{susapprox}) must obey the exact asymptotic
equations~(\ref{suslowomega}) and~(\ref{sushighomega}), whence:
\begin{eqnarray}
\label {tintcoffey}
\tint
&=&
\Deltauno/\eival_{1}
+
(1-\Deltauno)/\eivalW
\\
\tef^{-1}
&=&
\Deltauno\eival_{1}
+
(1-\Deltauno)\eivalW
,
\end{eqnarray}
These equations can be solved for $\Deltauno$ and $\eivalW$, yielding:
%
%
\begin{eqnarray}
\label{Deltaunoiden}
\Deltauno
&=&
\frac
{\tint \tef^{-1} - 1}
{\eival_{1} \tint - 2 + (\eival_{1} \tef)^{-1}}
\\
\label{LambdaWiden}
\eivalW
&=&
\frac 
{\eival_{1} - \tef^{-1}}
{\eival_{1} \tint -1}
\end{eqnarray}
This reduces the problem to the obtainment of the three characteristic
times $\eival_{1}^{-1}$, $\tint$, and $\tef$.
Naturally, they can be expressed in terms of the relaxation
eigenvalues:
%
\begin{equation}
\label{tintauto:tefauto}
\tint
=
\sum_{i}
\frac {\as_i}{\Lambda_i}
\;,
\qquad
\tef^{-1}
=
\sum_{i}
\as_i \Lambda_i
\;.
\end{equation}
However, the goal is to bypass the eigenvalue computation by
calculating them directly.


\subsection {Calculation of $\tint$, $\eival_1$ and $\tef$}
In the classical limit, Brown \cite{bro63} calculated the lowest
eigenvalue $\eival_{1}$ in the high barrier case (see also
\cite{aha69}).
Besides Coffey and co-workers derived $\tef$ \cite{cofcrekal93}.
On the other hand, Garanin calculated $\tint$ for a system of balance
equations to which Eq.~(\ref{BE}) can be reduced.
In this subsection we are going to extend the results for $\eival_1$
and $\tef$ to the quantum case (recovering the classical results in
the limit $S\to\infty$), and that of $\tint$ to the generic
system~(\ref{BE}) of balance equations.
This will result in a closed expression for the bimodal formula
(\ref{susapprox}).

\subsubsection {Integral relaxation time, $\tint$}

%
%
%
%
%
The calculation of $\tint$ is based in that the susceptibility can be
calculated analytically up to first order in $\Omega$ \cite{gar97}.
Exploiting this fact, one finds for $\tint$ (see App.~\ref{app:taus}):
\begin{equation}
\label{tintformula}
\tint
=
\frac{\beta}{\suseq}
\sum_{m=-S}^{S}
\frac{\Phi^2_m}{N_m^{(0)}P_{m+1|m}}
\end{equation}
%
%
%
where
\begin{equation}
\label{phim}
\Phi_m
=
\sum_{j=-S}^{m}
(\Mz-j)
N_j^{(0)}
.
\end{equation}
%
Notice that $\Phi_m$ only involves equilibrium averages, being
independent of the spin-bath coupling model.

The dependence of $P_{m \vl m'}$ in the denominator on the energy differences,
gives {\it minima} at the crossing fields \cite{fn:minima}
%
%
%
,
which remind of the {\it maxima} in the relaxation rate due to tunneling. 

\subsubsection{Lowest eigenvalue $\eival_1$}

The calculation of $\eival_1$, which corresponds to the Kramers rate,
constitutes an important task by itself.
Needless to say its relevance in the theory of activated processes
\cite{hantalbor90,mel91}.
In the classical case it is possible to derive an expression for
$\eival_1$ \cite{bro63}, giving good results for not too strong fields
\cite{aha69}.
In the quantum case, a closed analytical expression for $\eival_1$
exists only for $\hef = 0$ due to Villain and co-workers
\cite{viletal94} and later on reexamined and improved by W\"urger
\cite{wur98prl}.
Their result can be written in a form appropiate for future comparison
as:
%
%
\begin{equation}
\label{wurger}
\eival_{1}^{-1}
=
\sum_{m=0}^{S-1}
\frac 
{\theta_m}
{N_m^{(0)}P_{m+1|m}}
\end{equation}
with
$
\label{thetam}
\theta_m 
=
\sum_{j=m+1}^{S}
N_j^{(0)}
$.

\begin{figure}[!tbh]
\centerline{
\resizebox{9.cm}{!}{%
\includegraphics[angle=-90]{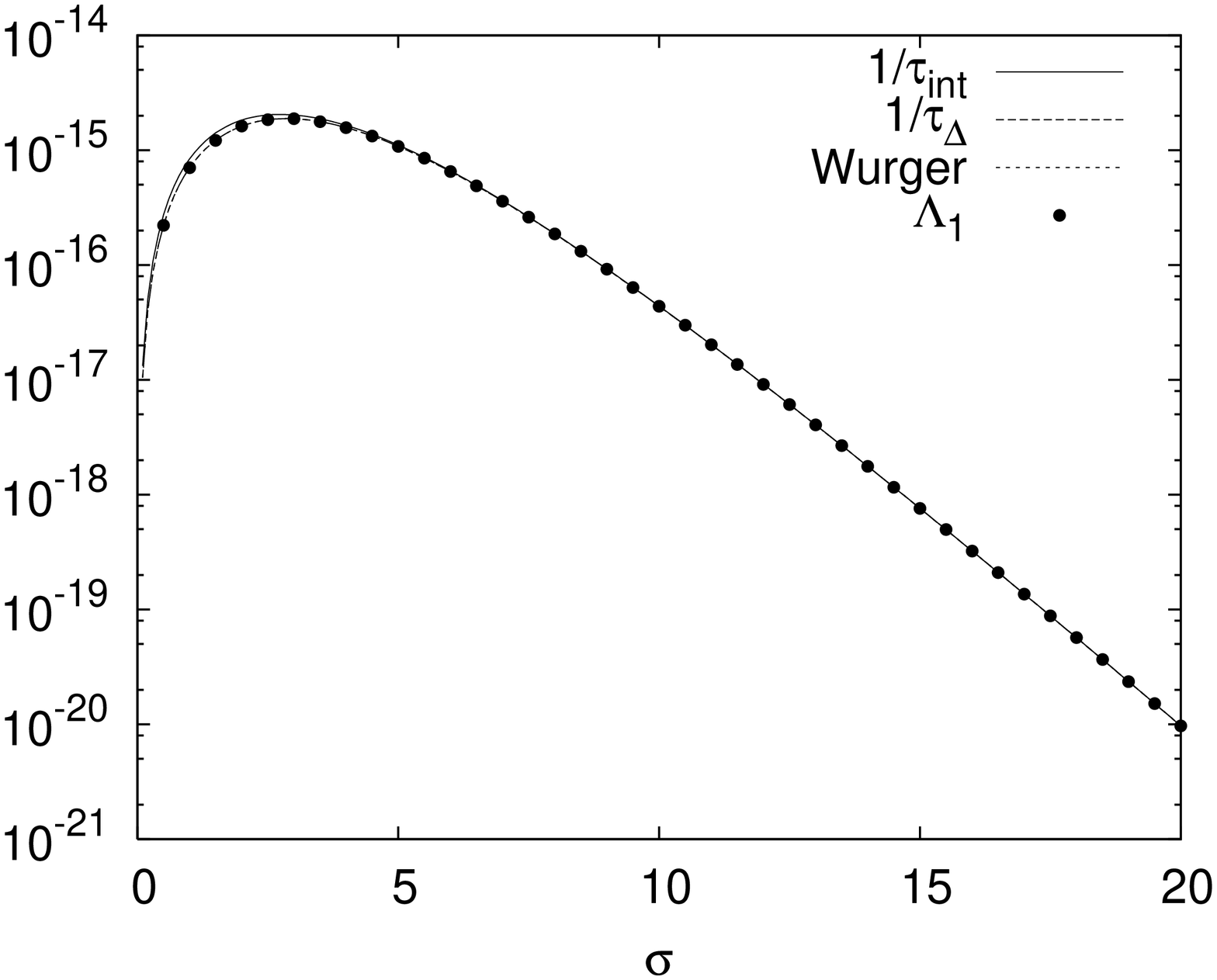}
}
}
\centerline{
\resizebox{9.cm}{!}{%
\includegraphics[angle=-90]{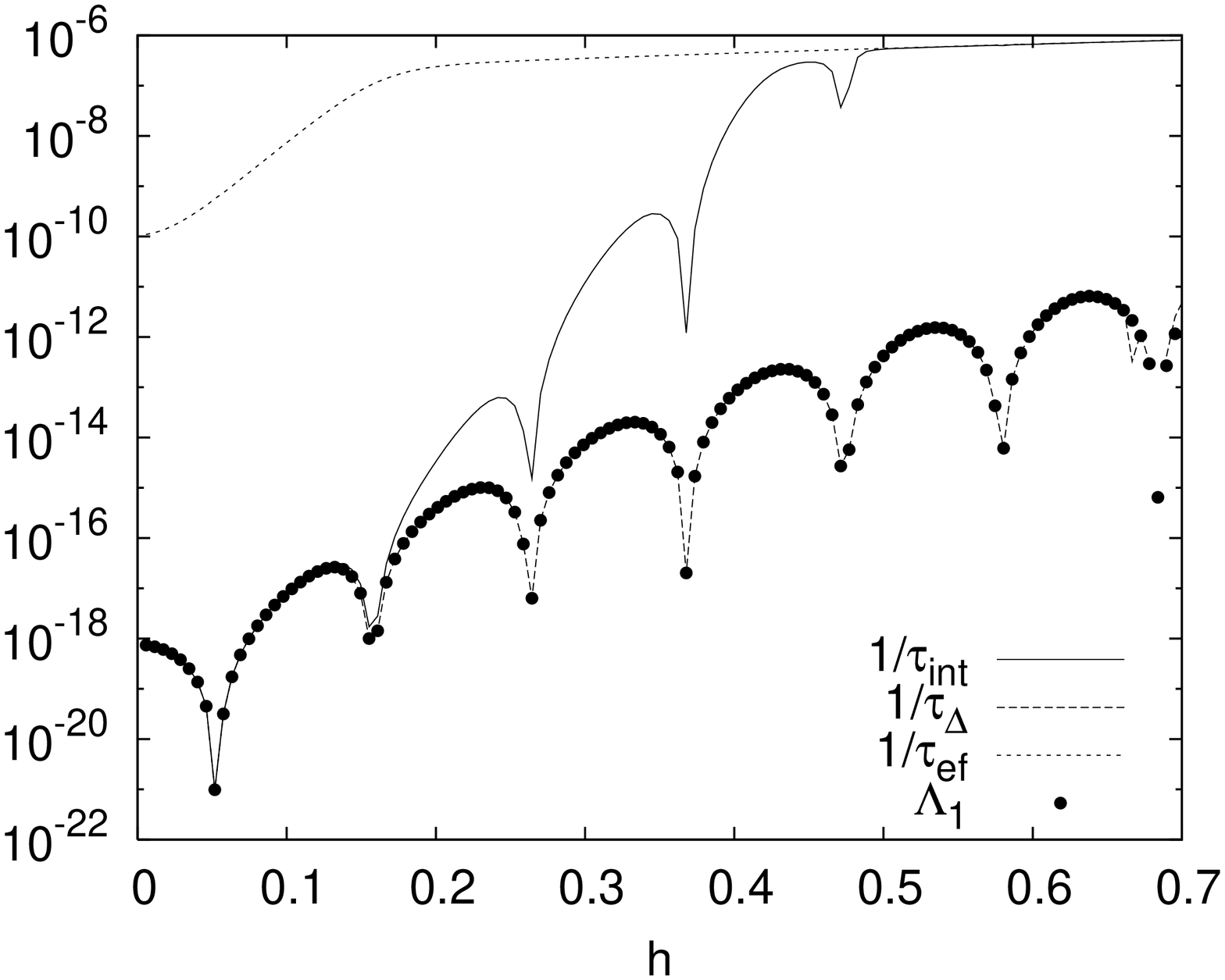}
}
}
\caption
{
Top: Comparison between the analytical $\tint ^{-1}$
    [Eq.~(\ref{tintformula})], $\tintN ^{-1}$ [Eq.~(\ref{wurger})] and the
    numerical lowest eigenvalue $\eival_{1}$ in the unbiased case
    as a function of $\sigma$, the barrier height.
The parameters are $S=10$, $T=0.1$ and $\lambda = 10^{-9}$.
    W\"urger's result and $\tintN$ have an maximum relative deviation
    of $10^{-3}$ in the plotted range (at the lowest barrier heights).
Bottom: $\tint^{-1}$, $\tintN^{-1}$, $\tef^{-1}$ [Eq.~(\ref{tef})] and
    $\eival_{1}$ for $S=10$, $\lambda = 10^{-9}$ and $\sigma=15$
    as a function of the field.
The relative difference between $\tintN^{-1}$ and $\eival_1$ never
exceeds the $1\%$ at not too strong fields ($\hef \leq 0.5$).
For $h \geq 0.7$ we find numerical instabilities at these values of $\sigma$
and $S$.
}
\label{fig:taus}
\end{figure}

Both $\eival_1$ from Eq.~(\ref{wurger}) and $\tint^{-1}$, together
with the exact numerical $\eival_1$ are displayed in
Fig.~\ref{fig:taus} showing their agreement at $\hef=0$.
%
%
However, as we increment the field $\tint^{-1}$ and $\eival_1$ deviate
from each other.
This is natural, should $\tint^{-1} \cong \eival_1$, the response
would be described by one mode, namely the overbarrier [i.e.,
$\Deltauno\cong 1$ in the bimodal approximation, see
Eq.~(\ref{tintcoffey})].
The same behavior was found in the classical limit
\cite{cofetal95,gar96}.
The physical reason for the disagreement between $\tint^{-1}$ and
$\eival_1$ is the intrawell processes entering into scene at
finite $\hef$.
Formally, at $h \ll 1$, the quantity $\phim^2/P_{m+1 \vl m}N_m^{(0)}$
is highly peaked at the barrier ($m = \mb$), and well approximated by
the overbarrier process; then $\tint^{-1} \cong \eival_1$.
Increasing the field, however, $\phim^2$ develops a second peak around
the bottom of the lower well, $m \cong S-1$, and the sum adquires a
relevant contribution from the intrawell processes \cite{gar96} (see also
Fig. \ref{fig:integrand}).
%
%
%
%

To generalise the calculation of $\eival_1$, we need an observable
where only overbarrier processes are reflected.
Then its associated integral relaxation time would probably be well
approximated by $\eival_1$.
This can be done generalizing to the quantum case the new integral
relaxation time, $\tintN$, introduced by Garanin for the classical
Fokker--Planck equation \cite{gar99}.
To this end, one chooses as observable the difference of the well
populations
$\Delta 
\equiv
N_{+}
-
N_{-}$.
To obtain the integral relaxation time associated to $\Delta$, we need
its low-$\Omega$ behavior, involving the ``static'' response $\chi _\Delta
\sim \partial_{B_z} \Delta$ [cf.\ the case of the
magnetization~(\ref{suslowomega})]:
%
\begin{equation}
\label{susDlowomega}
\chi_{\Delta}
(\Omega)
=
\frac
{\partial \Delta (\Omega)}
{\partial  \Bz}
\cong
\chi_{\Delta}
(1+
\iu
\Omega
\tintN
+
\dots
)
.
\end{equation}
%
This is done in App.~\ref{app:taus}, yielding the following $\tintN$
%
\begin{equation}
\label {tintN}
\tintN
=
\frac {\beta}{\chi_{\Delta}}
\sum_{m=-S}^{S-1}
\frac 
{\phim \phimN}
{N_m^{(0)}P_{m+1|m}}
\end{equation}
with ($\mb$ is the maximum level)
\begin{equation}
\label{phimN}
\phimN
=
\sum_{j=-S}^{m}
N_j^{(0)}
\big[
 \Delta - \sgn (j-\mb)
\big]
\end{equation}
%
It results that $\phim \phimN$ is nearly constant.
Then the sum in Eq.~(\ref{tintN}) has only one main contribution
(around the barrier $m \cong \mb$), not reflecting the dynamics inside
the wells, as it was intended.

We have seen in Fig.~\ref{fig:taus} that at $h=0$, $\tintN$ matches
W\"urger result~(\ref{wurger}) for $\eival_1$.
In fact, it can be seen that our $\tintN|_{\hef=0}$ reduces in the
high barrier regime to Eq.~(\ref{wurger}).  (see
App.~\ref{app:AFHBC}).
%
%
%
%
%
%
%
%
Another limit where we have an expression for $\eival_1$ is the
classical one.
Taking the limit $S \to \infty$ in $\tintN$ we indeed obtain Brown's
result in the high barrier range (App.~\ref{app:AFHBC})
%
\begin{eqnarray}
\label{Brown_result}
\nonumber
\eival_1 &\cong& 
\tau_D^{-1} \sigma^{3/2} \pi^{-1/2} (1 -\hef^2)
\big \{
(1+ \hef)
\exp [-\sigma (1 + \hef)^2]
\\ 
&+&
(1- \hef)
\exp [-\sigma (1 - \hef)^2]
\big \}
.
\end{eqnarray}
%
Finally, in the quantum case and $\hef \neq 0$, we have to check
$\tintN$ with the numerical results for $\eival_{1}^{-1}$; this was
done in Fig.~\ref{fig:taus}
%
%
%
%
%
There one sees that, contrary to the ordinary integral relaxation
time, $\tintN^{-1}$ provides a remarkable approximation for the
Kramers' rate.

\subsubsection{Effective time, $\tef$}

We conclude with the effective time.
By definition $\tef$ is the initial slope of the magnetization,
see Eq.~(\ref{sushighomega}).
The initial condition is the system at thermal equilibrium with $B_z =
B_z^{0} + \delta B_z$, so that $N_m(0)$ is known, whence
$\dot{N}_m(0)$ follows [directly from Eq.~(\ref{BE})].
Next, $\D \Delta M_z / \D t\, \vl_{t = 0}$ is calculated from
$\dot{N}_m(0)$ (see App.~\ref{app:taus} for details).
This gives the following expression for $\tef$:
\begin{equation}
\label{tef}
\tef^{-1}
=
\frac {\beta}{\suseq}
\sum_{m = -S}^{S}
N_m^{(0)}P_{m+1|m}
.
\end{equation}

The classical limit of Eqs.~(\ref{tintformula}), (\ref{tintN})
and~(\ref{tef}) give the expressions derived over the years from the
Fokker--Planck equation~(\ref{f-pclassical}).
%
They are the
sought analytical expressions for the characteristic times $\tint$,
$\eival_1^{-1}$ and $\tef$ in the quantum case.
Together with Eqs.~(\ref{Deltaunoiden}) and~(\ref{LambdaWiden}) they
provide a closed formula for the response under the bimodal
approximation, which can be checked against exact results.


\section{Analysis of the dynamic response}
\label{sec:R}

In this section we study the full dynamical response.
We compare exact numerical results with the approximate bimodal
formula constructed in the previous section.
We also present approximate tractable expressions.


\subsection{Exact results vs.\ bimodal formula}

\begin{figure}[!tbh]
\centerline{
\resizebox{9.cm}{!}{%
\includegraphics[angle=-90]{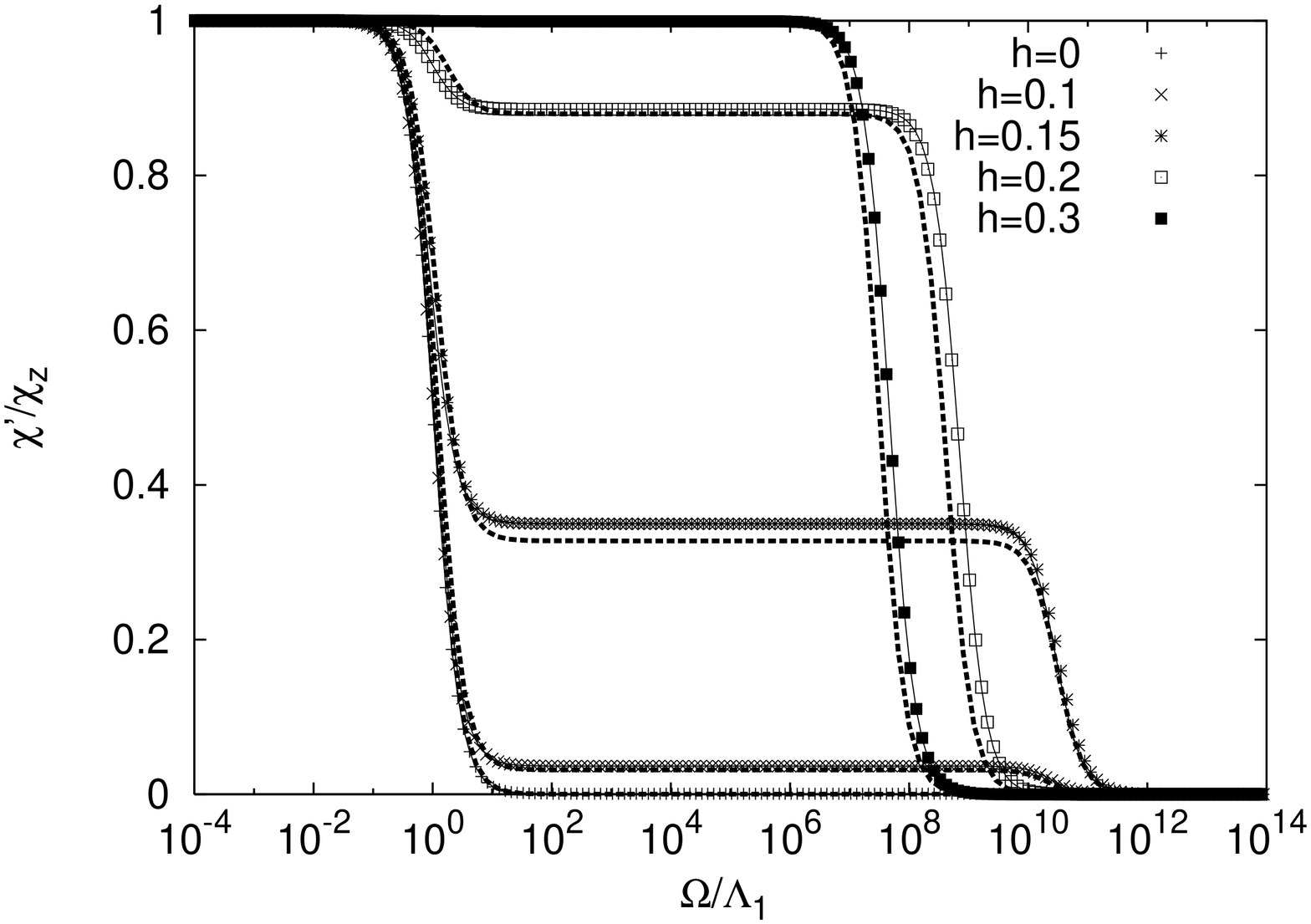}
}
}
\centerline{
\resizebox{9.cm}{!}{%
\includegraphics[angle=-90]{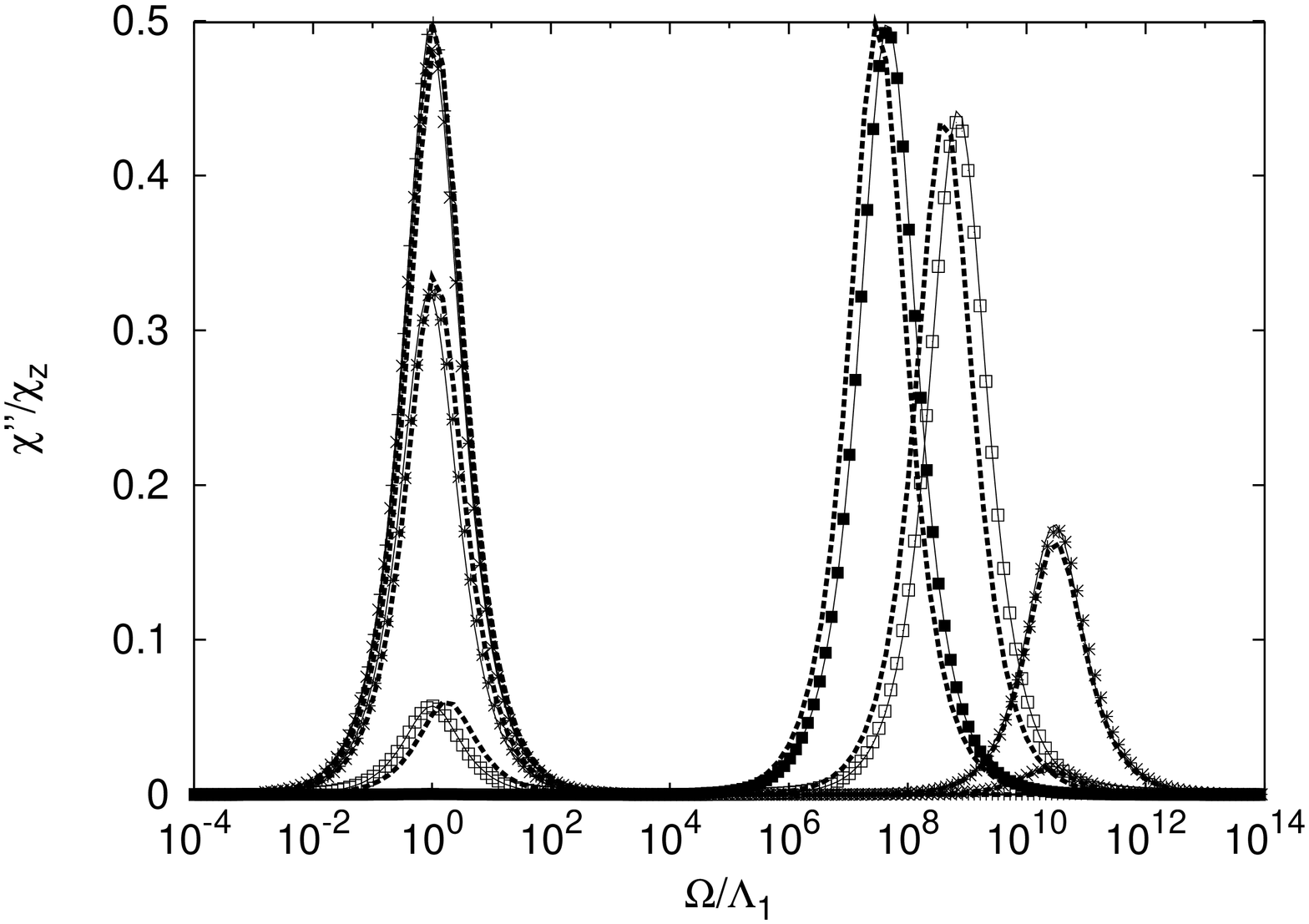}
}
}
\caption{
Real parts (top) and imaginary parts (bottom) of the dynamical
susceptibility of a spin $S=10$ with $\lambda = 10^{-9}$, at $T=0.1$
and $\sigma = 15$ in various applied fields.
The symbols represent the exact numerical results and the solid
lines the bimodal formula~(\ref{susapprox}).
 The dashed (thick) lines correspond to the bimodal formula but
  with the approximate $\Deltauno$, $\eival_{1}$, and $\eivalW$ from
  Eqs.~(\ref{Deltaunohighsigma}), (\ref{Arrhenius}),
  and~(\ref{eival_1/eival_W}).  
}
\label{fig:susS10}
\end{figure}

\begin{figure}[]
\centerline{
\resizebox{9.cm}{!}{%
\includegraphics[angle=-90]{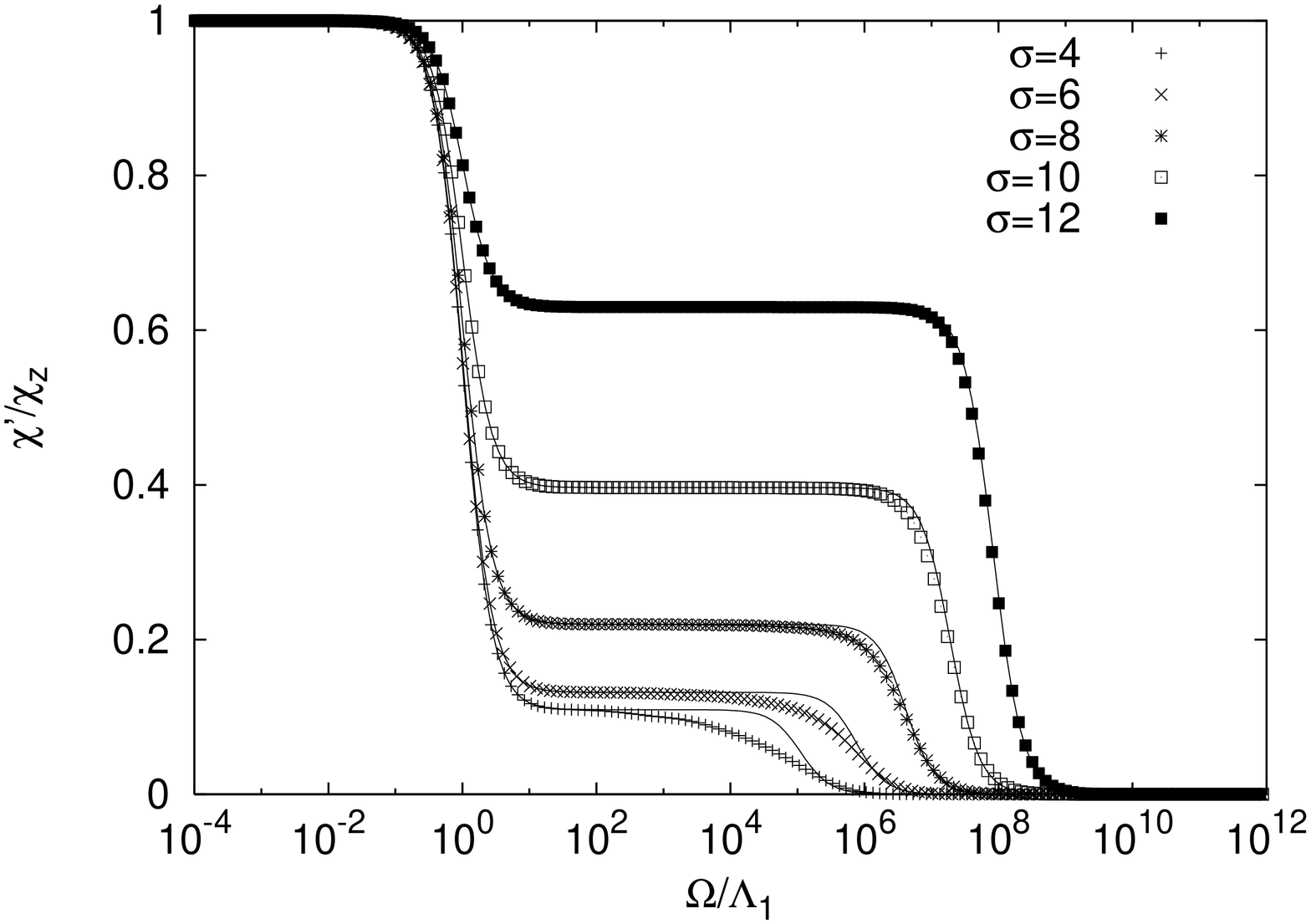}
}
}
\centerline{
\resizebox{9.cm}{!}{%
\includegraphics[angle=-90]{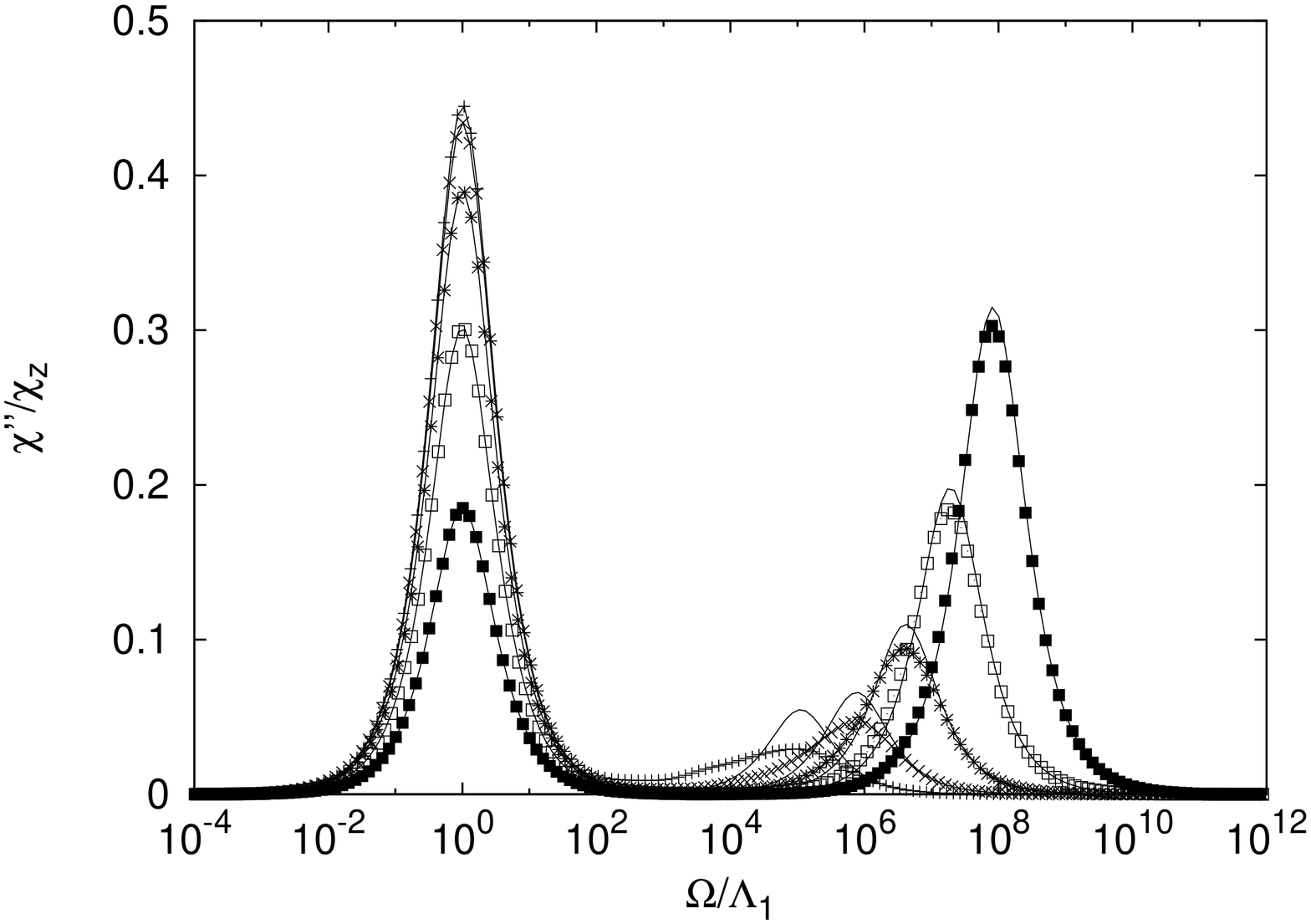}
}
}
\caption {
Real parts (top) and imaginary parts (bottom) of the dynamical
susceptibility of a spin $S=10$ with $\lambda=10^{-9}$ at $T=0.1$ and
$\hef=0.2$.
Here $\sigma$ = 4, 6, 8, 10 and 12.
Points represent the exact numerical results and lines are the
theoretical formula~(\ref{susapprox}).
}
\label{fig:susS10_hef0p21}
\end{figure}

To obtain numerically exact results for the response [see
Eqs.~(\ref{response1}) and~(\ref{dynamicalsus})], we compute
numerically the eigenvalues $\eival_i$ and the amplitudes $\as_i$ [see
Eq.~(\ref{c_i})].
%
%
%
Comparison between them and the analytical formula~(\ref{susapprox})
is shown in Figs.~\ref{fig:susS10} and~\ref{fig:susS10_hef0p21}.
%
%
In Fig.~\ref{fig:susS10} we plot the field dependence of the
susceptibility spectra.
The agreement is excellent in the low temperature range (large
$\sigma$).
%
The $\sigma$-dependence is presented in Fig.~\ref{fig:susS10_hef0p21}.
%
%
The slow dynamics at $\Omega \sim \eival_1$ is well described by the
analytical expression, while reducing sufficiently $\sigma$ the
bimodal approximation starts to fail at high frequencies
$\Omega\sim\eivalW$.
Specifically, the second peak in the exact imaginary part is no longer
a single Lorentzian and a broadening with respect to the analytical
curve is observed.

The reason for the overall agreement between the bimodal approximation
and the exact results can be understood as follows.
The overbarrier dynamics must be well captured, since it is exactly
given by $\eival_1$, while $\eival_1$ is very well approximated by
$\tintN$ (see Sec.~\ref{sec:AA}).
Next, in the two mode approximation we assumed that the intrawell
modes are close to one another, so that they can be described with the
collective scale $\eivalW$.
We have checked that the degree of closeness does not depend on
$\sigma$ so as to explain the disagreement for low $\sigma$ and high
$\Omega$ in Fig.~\ref{fig:susS10_hef0p21}.
%
%
What happens is that at high $\sigma$ the bimodal approximation works
very well because only intrawell processes involving the lowest well
states are accessible, yielding $\as_i \sim 0$ for the rest of the
modes.
For the example considered, $S=10$, one indeed sees that at
$\sigma=15$, only $\as_{17}$ and $\as_{18}$ contribute (involving $m =
\pm S$ and $m \pm (S-1)$), and to a smaller extent $\as_{16}$ and
$\as_{19}$ (see the inset in Fig.~\ref{fig:eigenvalues}).
%
%
Lowering enough the barrier more and more intrawell modes play a role
yielding a broadening of the curve for the imaginary part of $\sus$.
However, we have to go down to $\sigma=6$ to find large disagreements,
while most experiments in these systems involve $\sigma\gtrsim10$--$20$.


\subsection{Intrawell versus overbarrier processes}
\label{subsec:IVO}

\begin{figure}[h!]
\centerline{
\resizebox{8.cm}{!}{%
\includegraphics[angle=-90]{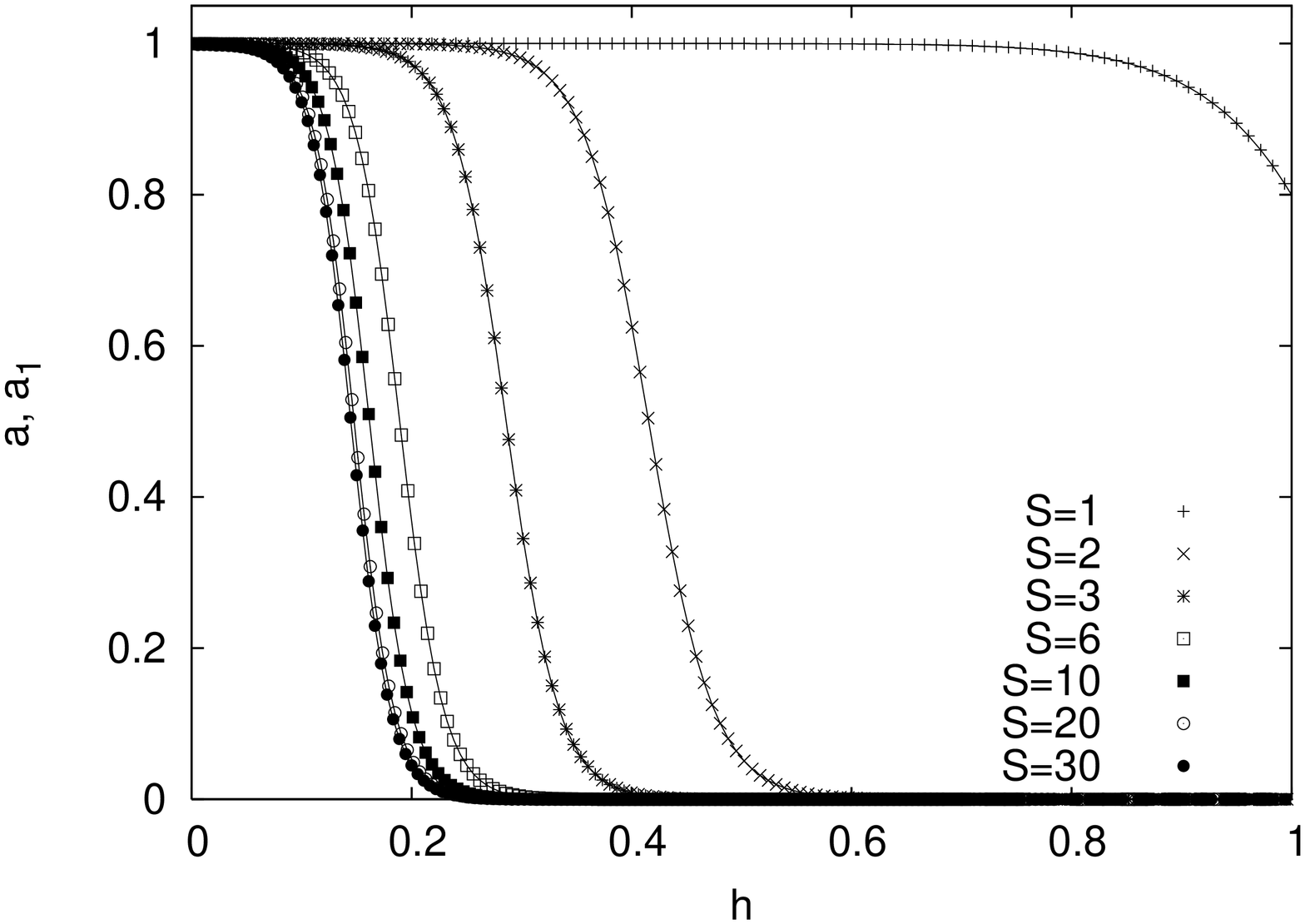}
}
}
\centerline{
\resizebox{8.cm}{!}{%
\includegraphics[angle=-90]{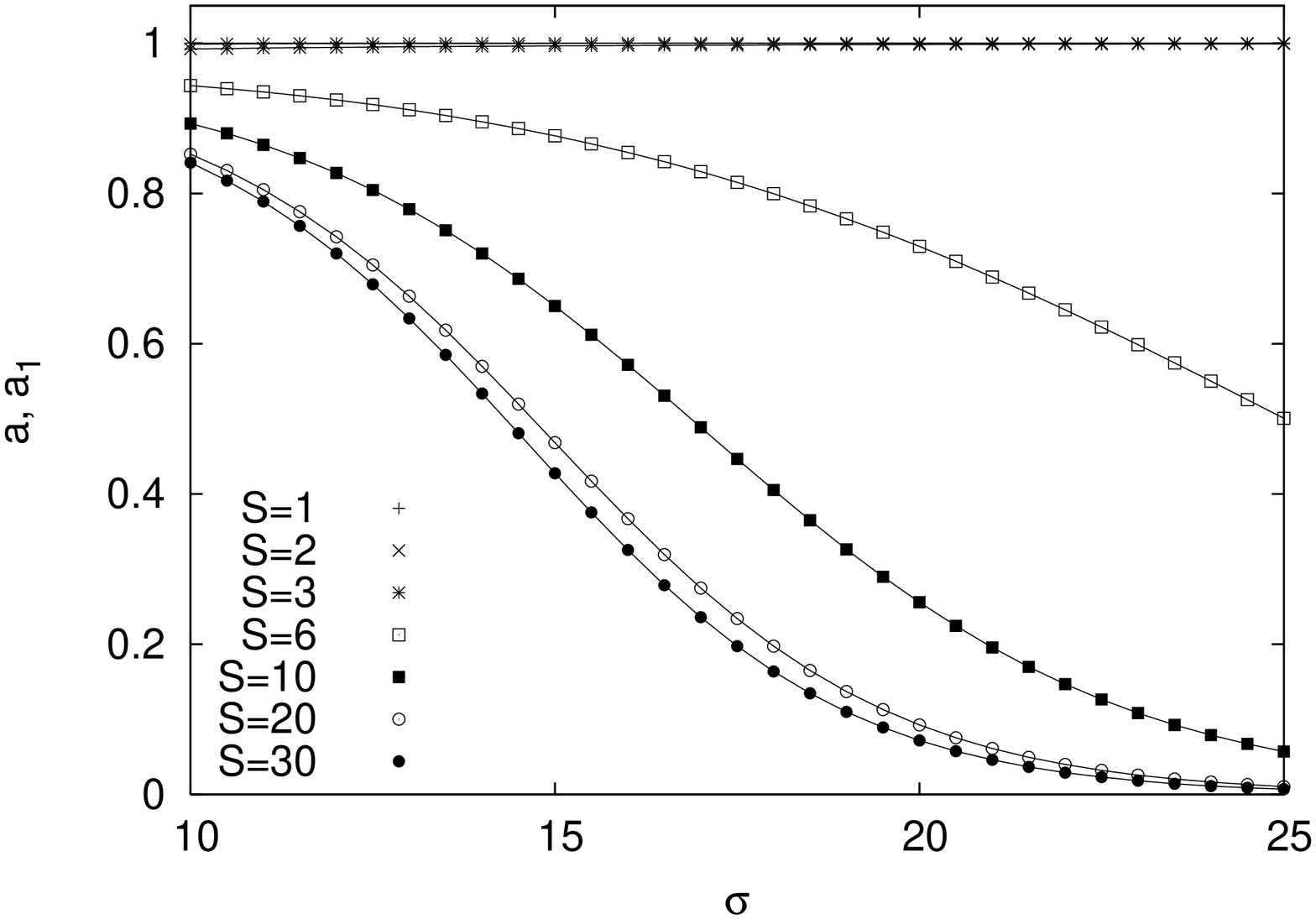}
}
}
\centerline{
\resizebox{8.cm}{!}{%
\includegraphics[angle=-90]{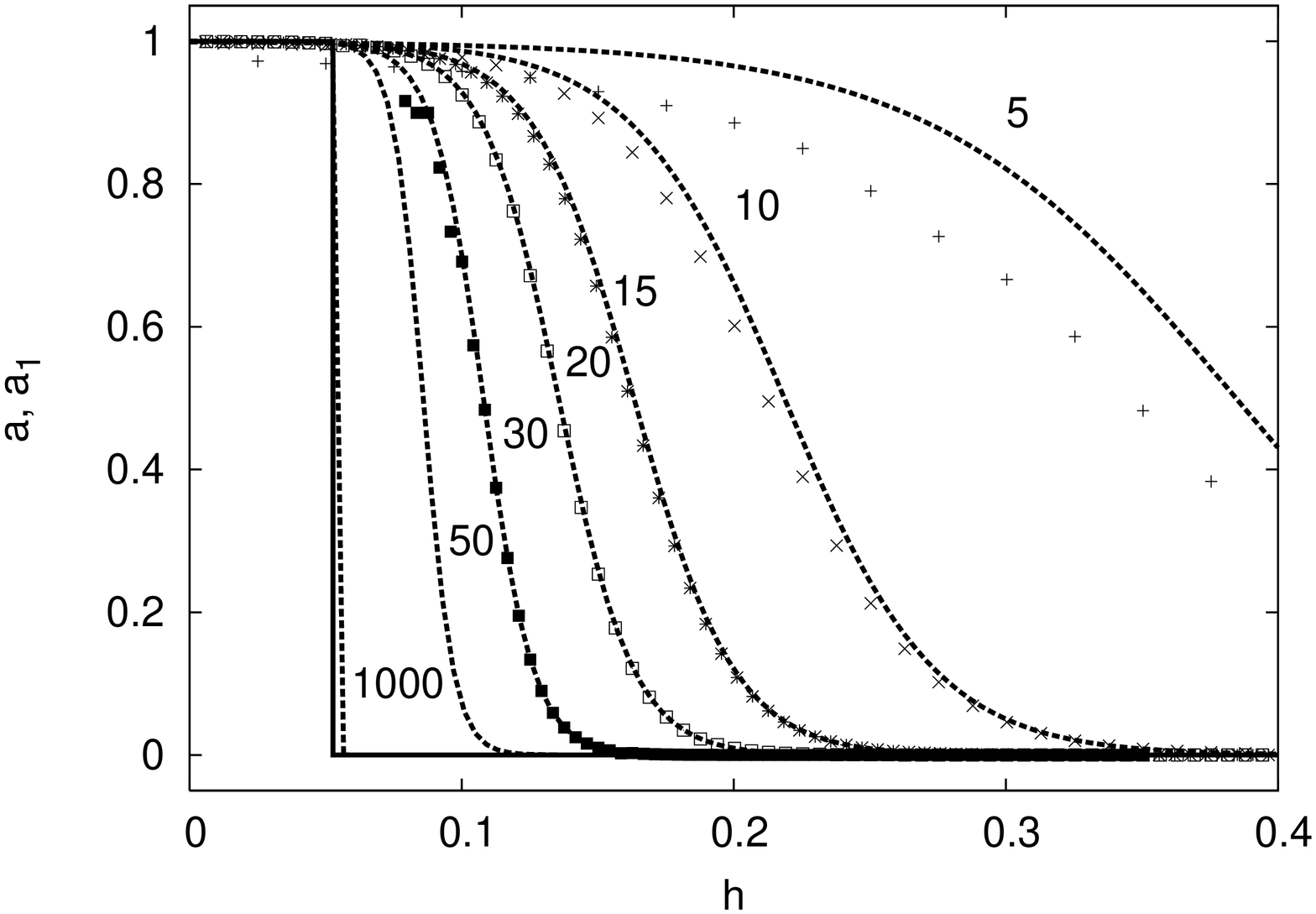}
}
}
\caption{
Top: Field dependence of $\as_1$ (points) and $\Deltauno$ (lines) for
different values of $S$.
We have fixed $\sigma = 15$, $\lambda = 10^{-9}$ and $T = 0.1$.
Middle: Barrier dependence of $\as_1$ and $\Deltauno$ at the same
$S$'s.
The effective field has been fixed at $\hef = 0.15$.
Bottom: Field dependence of $\as_1$ and the approximate $\Deltauno$ of
Eq.~(\ref{Deltaunohighsigma}).
The numbers stand for the $\sigma$ value.
Increasing $\sigma$ the curves approach a step function (solid line).
At $\sigma = 30$ we find numerical instabilities so that for
$\sigma=50,1000$ only the approximate formula is plotted.
}
\label{fig:A-h-sigma}
\end{figure}

Let us study the relative importance of intrawell and overbarrier
dynamics in the response, a subject that recieved some attention in
the classical case \cite{cofetal95, gar96}.
%
%
%
%
The competition between them becomes mediated by $\Deltauno$ in our
Eq.~(\ref{susapprox}).
In particular when $\Deltauno = 1$ only relaxation across the barrier
takes place while at $\Deltauno = 0$ the opposite occurs.
From Fig.~\ref{fig:susS10} and~\ref{fig:susS10_hef0p21}, we see that
the relative weight of the two modes depends both on $\hef$ and the
barrier height.
The field and $\sigma$ dependence of $\Deltauno$ is drawn in
Fig.~\ref{fig:A-h-sigma} for different $S$.
The main features are as follows.
Concerning the $\hef$ dependence, the transition from overbarrier
dominated ($\Deltauno\sim1$) to intrawell dominated response
($\Deltauno\sim0$) is produced in a relatively small window of $\hef$.
Secondly the transition occurs at larger values of $\hef$ the smaller
the value of $S$ is.
In addition, the $\sigma$ dependence of $\Deltauno$ shows that the
value of $S$ is also crucial in the limit $\sigma \to \infty$.
In this limit $\Deltauno$ tends to $1$ or to $0$ depending on $S$.
%
%
%
%
%

Let us try to explain qualitatively these results, invoking a
statistical mechanical argument.
Inspecting the level structure of the spin Hamiltonian, we see that at
$\xi=\sigma /S$ the levels $-m$ and $m-1$ become degenerated.
Such field corresponds to 
%
\begin{equation}
\label{hcritical}
\hcr \equiv
\frac {1}{2 S - 1}
.
\end{equation}
Then, if $\hef < \hcr$ the first excited level is $m = -S$ (in the
other well), whereas if $\hef > \hcr$ it is $m = S-1$ (inside the same
well).
Approaching the large $\sigma$ limit we just need to consider the
fundamental state and the first excited level.
Then, when $\sigma \to \infty$, if $\hef < \hcr$ intrawell process are
inhibited, because only $m = S$ and $m = -S$ are populated, giving
$\Deltauno = 1$.
On the contrary if $\hef > \hcr$, the states that contribute are $m =
-S$ and $m = S-1$, in the same well, so that $\Deltauno = 0$.

The above argument also serves to explain the large $\sigma$ limits in
Fig.~\ref{fig:A-h-sigma}, because the field used, $\hef = 0.15$, is
lower than $\hcr (S = 1,2,3)$, and greater than $\hcr (S >3)$.
On the other hand, we have $\hcr \to 0$ when $S \to \infty$.
This is in agreement with the classical calculation where $\Deltauno
\to 0$ as soon as one makes $\hef \neq 0$ in the limit $\sigma \to
\infty$ \cite{gar96}.
Eventually our statistical mechanical argument also explains the $S$
dependence of the onset of intrawell modes when increasing the field
(Fig.~\ref{fig:A-h-sigma}).
Specifically, it gives a step function $\Theta (\hcr - \hef)$ at
$\sigma \to \infty$, while the remaining curves can be understood as a
smoothing of the abrupt step due to finite temperature
(Fig.~\ref{fig:A-h-sigma}).

%
%
%
%
%


\subsection{Approximate expressions in the Kramers regime}
\label{sec:QATTE}

To conclude we present approximate formulas for the two time scales of
the response, $\eival_1$ and $\eivalW$, in the high barrier (Kramers)
regime, and for the parameter $\Deltauno$ controlling their relative
weight in the response.
In this regime $\Deltauno$ is well described by the formula (see
App.~\ref{app:AFHBC}):
%
\begin{eqnarray}
\label{Deltaunohighsigma}
\Deltauno
\cong
\big[1
+ 
\tfrac {1 }{4}
\,
(1- \factorq)^2
\,
\e^{\sigma ( \factorq + 1 ) ^2 ( \hef - \hcr )}
\big]^{-1}
\end{eqnarray}
%
where $q=(S-1)/S$.
Returning for a moment to the previous discussion, note that this
formula captures the $\hef$, $\sigma$ and $S$ dependences of
Fig.~\ref{fig:A-h-sigma};
for instance it approaches the step function $\Theta(\hef-\hcr)$ when
$\sigma \to \infty$, and gives its smoothening for finite $\sigma$.
Note finally that Eq.~(\ref{Deltaunohighsigma}) depends only on
parameters of the spin Hamiltonian, being independent of the coupling
(as in the classical case \cite{gar96}).

%
%
%
%
Let us go now into the Kramers rate.
At low temperatures $\eival_1$ can be expressed in terms of the
barriers $\Delta U_{\pm}\equiv\varepsilon_{\mb}-\varepsilon_{\pm S}$
in a clear Arrhenius form (App.~\ref{app:AFHBC})
\begin{equation}
\label{Arrhenius}
\eival_1 
\cong 
\left (
\e^{-\beta \Delta U_+}
+
\e^{-\beta \Delta U_-}
\right )
\Big(
\frac {1}{P_{\mb +1 \vl \mb}}
+
\frac {1}{P_{\mb -1 \vl \mb}}
\Big)^{-1}
.
\end{equation}
Recall that the transition probabilities
$ P_{m \vl m'} = \vl L_{m, m'} \vl ^2 W_{m \vl m'} $
[Eq.~(\ref{identification})], involve the coupling matrix elements
$ L_{m, m \pm 1} = (2m \pm 1) \sqrt {S(S+1) - m(m \pm 1)}$
and the rates
$\Wu_{m\vl m'}
=
\lambda\,\tf_{m\,m'}^{\kk}/(\e^{ \beta \tf_{m\,m'}} -1)$.
%
%
Equation~(\ref{Arrhenius}) generalizes the zero-field result of
Villain and co-workers \cite{viletal94}.
In addition, with the appropriate $L_{m, m \pm 1}$ and the spectral
index $\kk$ (from $J(\omega)=\lambda\, \omega^{\kk}$), it is
applicable to other couplings and baths.

Finally we turn to the fastest mode $\eivalW$.
In order to fix some time scale in the problem, we measure $\eivalW$
relative to $\eival_1$.
Under the same assumptions used to obtain the above $\Deltauno$ and
$\eival_1$, we find for $\eivalW/\eival_1$:
\begin{eqnarray}
\label{eival_1/eival_W}
\frac {\eivalW}{\eival_1}
&=&
\frac {1}{4 S^2}
\frac {\Deltauno}{1 - \Deltauno}
\\
\times
&\bigg \{&
\e^{\beta \Delta U_+}
\Big [
P_{S-1 \vl S}
\Big (
\frac {1}{P_{\mb +1 \vl \mb}}
+
\frac {1}{P_{\mb -1 \vl \mb}}
\Big )
\Big ]
\nonumber\\ 
& &
{}+
\e^{\beta \Delta U_-}
\Big [
P_{-S+1 \vl S}
\Big (
\frac {1}{P_{\mb +1 \vl \mb}}
+
\frac {1}{P_{\mb -1 \vl \mb}}
\Big )
\Big ]
\bigg \}
\nonumber
\end{eqnarray}
%
While the parts involving the barriers $\Delta U_{\pm}$ and
$\Deltauno$, are independent of the coupling model, the transition
rates at the wells and at the barrier are quite sensitive to the
interaction.
Thus the ratio $\eivalW/\eival_1$ can be used to compare different
couplings and baths.

Formulas~(\ref{Deltaunohighsigma}), (\ref{Arrhenius})
and~(\ref{eival_1/eival_W}) provide tractable expressions appropiate
for the experiments in the superparamagnetic range, out of resonance
conditions (see App. \ref{app:AFHBC} for further approximate expressions).
 Their reasonable agreement with the
  exact numerical results can be seen in Fig.~\ref{fig:susS10}.
Under resonant conditions, we expect that our equations will still
work after some modifications.
For example, the barrier height will be lowered, since $\mb$ would not
be the barrier top but the state at which the activated tunnel process
takes place.

\section {Summary}
\label{sec:C}

The relaxation theory of a quantum superparamagnet has been developed.
We have studied the minimal model, namely a spin of arbitrary $S$ with
uniaxial anisotropy in a longitudinal field.
Scarce results were available for the full dynamical response in the
presence of external fields.
Still, the popular single-Debye form for the susceptibility spectra
could be expected to fail, in analogy with the classical situation.
Here, these topics have been addressed in the frame of quantum balance
equations rigorously grounded on the system-plus-bath approach to
quantum dissipative systems.

We began analyzing the eigenstructure of the relaxation matrix
associated to the system of balance equations.
The form of the eigenvectors allowed us to identify and classify the
different relaxation mechanisms.
In this way, full content has been given to popular statements about
the relation of eigenvalues and relaxation modes.
Besides, it has been put in connection with the Sturm--Liouville
eigenstructure of the classical Fokker--Planck limit (parity, nodes,
etc.).

Two main processes emerge: activation over the anisotropy barrier,
with a time scale  $\eival_1^{-1}$, and a bunch of close (in
logarithmic scale) fast intrawell modes, with an ``average'' time
scale $\eivalW^{-1}$.
The identification and separation of the modes suggest the
introduction of an approximate two-mode expression for the dynamical
susceptibility.
Then, following the approach of Kalmykov {\em et al.} for the
classical case, the parameters of that formula are expressed in terms
of three characteristic quantities: integral relaxation time $\tint$,
lowest non-vanishing eigenvalue $\eival_1$ (Kramers rate), and the
initial slope of the magnetization decay $\tef$.
The usefulness of the classical approach was that these three times
could be obtained analytically (which was done over the years for
different relevant situations).
Then, we were left with the task of finding them all in analytical
form in the quantum domain.
This has been accomplished here, getting formulas for the quantum
$\tint$, $\eival_1$ (via the integral relaxation time for the
population difference $\tintN$) , and $\tef$, which recover known
classical results when taking the $S\to\infty$ limit.
With them, one has a bimodal equation for the dynamical susceptibility
of a spin with arbitrary $S$ where all ingredients can be expressed in
closed form.

We have compared exact numerical results with such bimodal expression;
the formula results to be quite accurate, specially in the
superparamagnetic regime ($\Delta U/\kT \sim 10$--$20$; the range of
major experimental interest).
Its limits of validity have been assessed and interpreted in physical
terms.
Furthermore, in the range where the bimodal description works best, we
have derived simple analytical expressions for $\eival_1$, $\eivalW$
and $\Deltauno$ (the parameter controlling the relative weight of the
two effective modes).
These generalize formulas available at zero field, while they can also
be applied to other structures of the coupling and spectral densities
of the bath (e.g., Kondo coupling to electron-hole excitations).

In the presence of terms not conmuting with $S_z$ in the Hamiltonian
(e.g., transverse fields) the situation is expected to be altered by
tunnel events between resonant levels $-m$ and $m-k$.
Then, the balance equations loose the three-term recurrence form which
allowed to derive the closed-form solutions \cite{leulos00}.
However, in the superparamagnetic range, thermal activation is still
expected to govern the physics.
Actually, in non-resonant conditions (the generic case), tunnel is
inhibited, while at the resonances, it can be accounted for
heuristically by lowering the effective barrier a few states.
Thus, we hope that the two-mode picture will be of use with the
appropriate amendments.

The formulas derived cover experiments even under moderate and strong
fields, where few work had been done.
The onset of competence of the dynamics inside the wells with the
overbarrier mode would occur at fields of a few Tesla, so that
comparison with experimental data would be possible.
We finally remark that the treatment employed covers from the deep
quantum case ($S \sim 1$) to the classical regime ($S\gg1$).
Thus, the equations derived recover, in the limit $S\to\infty$, the
classical work on magnetic nanoparticles.


\begin{acknowledgments}
  This work was supported by DGES, project BFM2002-00113, and DGA, project
  {\sc pronanomag} and grant no.~B059/2003.
  We acknowledge F. Luis, L. Mart\'{\i}n-Moreno and D. Prada for useful
  discussions and their support during this
  work.
\end{acknowledgments}


\appendix


\section {CLASSICAL LIMIT OF BALANCE EQUATIONS}
\label{app:CL}
In this appendix we consider the classical limit of the balance
equation~(\ref{BE}).
%
%
Any limit procedure requires to specify which quantities are kept
constant and which scaled variables are needed to monitor the
evolution.
For the classical limit, a natural choice is to maintain fixed
$\sigma$ and $\xi$.
At constant $T$ this implies keeping the anisotropy-barrier height and
amount of Zeeman energy constant.
%
%
This means, that the levels tend to a continuum, so 
$\tf_{m,m \pm 1} \to 0$.
Then we need the
behavior of
$P_{m \vl m \pm 1}$  near  $\tf_{m, m \pm 1}= 0$.   
Expanding $\Pmmprima$
up to first order, with 
$d_x f(x)|_{x=0} = - d_x
f(-x)|_{x=0}$, and the detailed balance condition
(\ref{detailed}):
\begin{equation*}
\frac {\D \Pmmprima} {\D \tf_{m\,m'}}
 \Big |_{\tf_{m\, m'} = 0}
=
-
\beta
\frac{1}{2}
\Pmmprima ( \tf_{m\, m'} = 0)
,
\end{equation*}
therefore,
\begin{equation}
\label{expansionPs}
\Pmmprima =
\Pmmprima (0)
-
\frac{\Pmmprima (0)}{2}
\beta \tf_{m\,m'}
+
\orden (\beta\tf_{m\,m'}^2) 
\end{equation}
%
Now, it results convenient to introduce the notation,
\begin{equation}
\label{notation}
\p_m \equiv \Pmplusm (0)
;
\qquad
\p_{m-1} \equiv \Pmmminus (0)
,
\end{equation}
identifying through this appendix $\tf_{m\,m'} \equiv \beta
\tf_{m\,m'}$
and inserting~(\ref{expansionPs}) in Eq.~(\ref{BE}),
\begin{eqnarray}
\label{limit_eqprevious}
\dot 
N_m
&=&
\big (
\p_m + \frac{\p_m}{2}
\tf_{m+1,m}
\big )N_{m+1}
\nonumber\\
&-&
\big (
\p_m - \frac{\p_m}{2}
\tf_{m+1,m}
\big )N_{m}
\nonumber\\
&+&
\big (
\p_{m-1} - \frac{\p_{m-1}}{2}
\tf_{m,m-1}
\big )N_{m-1}
\nonumber\\
&-&
\big (
\p_{m-1} + \frac{\p_{m-1}}{2}
\tf_{m,m-1}
\big )N_{m}
\end{eqnarray}
where we have utilized~(\ref{notation}). Now, writing $\p_{m-1} = \p_m +
[\p_{m-1} - \p_m]$ and multiplying the right hand side of
(\ref{limit_eqprevious}) by $S^2/S^2$ we obtain,
\begin{widetext}
\begin{eqnarray}
\label{limit_eq}
\nonumber
\dot N_m
&=&
\frac{\p_m}{S^2}
\Big [S^2
(N_{m+1} - 2N_m + N_{m-1})
\Big ]
\\ \nonumber
&+&
\frac{\p_m}{S^2}
\Big [
S^2
\frac{1}{2}(
\tf_{m+1,m}N_{m+1}
+
\tf_{m+1,m}N_{m}
-
\tf_{m,m-1}N_{m}
-
\tf_{m,m-1}N_{m-1}
)
\Big ]
\\ 
&+&
S
\bigg ( \frac {\p_m}{S^2} -\frac {\p_{m-1}}{S^2} \bigg )
\Big [
S(N_{m} - N_{m-1} )+
\frac{S}{2}
\tf_{m,m-1} (N_m + N_{m-1})
\Big ]
\end{eqnarray}
\end{widetext}
%
%
%
In the first line of (\ref{limit_eq}) we recognize the usual
discretization of the second derivative 
\begin{equation}
\label{2derivative}
S^2
(N_{m+1} - 2N_m + N_{m-1})
\rightarrow 
\frac{\partial^2 \W (z,t)}{\partial z^2}
\end{equation}
where $\W (z,t)$ is the classical spin distribution.
The third line can be identified with the
discretizations of the first derivatives:
\begin{eqnarray}
\label{1derivative}
\nonumber
S(N_{m} - N_{m-1} )
&\rightarrow&
\frac{\partial \W (z,t)}{\partial z}
\\ \nonumber
S\tf_{m+1,m}= \beta S \big ( \varepsilon_{m+1} - \varepsilon_{m})
&\rightarrow &
\frac{\partial u (z)}{\partial z}
\\ 
S
\bigg ( \frac {\p_{m+1}}{S^2} -\frac {\p_m}{S^2} \bigg )
&\rightarrow&
\frac{\partial D(z)}{\partial z}
\end{eqnarray}
here, we have utilized that $\p_m/S^2 \rightarrow D(z)$.
The function $D(z)$ depends on the specific form of $\p_{m} (\equiv
\Pmplusm (0))$ and $u(z)$ is the classical energy function [$u(z)
\equiv \beta \Hs(z) $].
Finally, the second line is the discretization
of $\frac {\partial}{\partial z} \Big (\frac {\partial
u(z)}{\partial z} \Big ) \W(z,t)$, since
\begin{eqnarray}
\label{mixedderivative}
\nonumber
S^2
\frac{1}{2}&(&
\tf_{m+1,m}N_{m+1}
+
\tf_{m+1,m}N_{m}
-
\tf_{m,m-1}N_{m}
\\
&-&
\tf_{m,m-1}N_{m-1}
)
 \rightarrow
\frac {\partial}{\partial z}
\Big (
\frac {u(z)}{\partial z}
\Big )
\W(z,t)
\end{eqnarray}
%
Using (\ref{2derivative}), (\ref{1derivative}) and
(\ref{mixedderivative}), we obtain the continuum version of
(\ref{BE}), Eq.~(\ref{f-pclassical}), where $D(z)\equiv P_{m+1 \vl
m}(0)$

Let us emphasize that this derivation only makes use of detailed
balance condition~(\ref{detailed}), and a nonvanishing transition
probability at zero frequency $P_{m+1 \vl m}(0)$, which has been
tacitly assumed from the beginning.
%
%
Considering interaction to electron-hole excitations ($F \propto
S_{\pm}$), with an spectral density $J(\omega) \propto \omega$ then
$D(z)\propto(1 - z^2)$ [see Eqs.~(\ref{identification}) and~(\ref{W})
in Sec.~\ref{sec:QME_A}].
On the other hand, coupling to phonons (including two phonon
processes) yields $D(z) \propto z^2 (1 - z^2)$.

\section {DETAILS FOR THE CALCULATION OF $\tint$, $\tintN$ and $\tef$}
\label{app:taus}

Here we calculate the three time constants,
$\tint$, $\tintN$ and  $\tef$, which characterize the magnetization relaxation
in the bimodal approximation [see Sec.~\ref{sec:AA}].


{\it (i) $\tint$}

As $\tint$ describes the low-$\Omega$ behavior of the
susceptibility our purpose here is to obtain $\sus$ up to first order in
$\Omega$ [see Eq.~(\ref{suslowomega})].
For future convenience, we write $N_m$  as,
\begin{equation}
\label{Nmrewrite}
N_m \cong N_m^{(0)} + N_m^{(1)} = 
N_m^{(0)} 
\big (
1 + \beta q_m
\big )
\end{equation}
%
where $N_m^{(0)}$ and $N_m^{(1)}$ are the zero and first order (in $\delta
B_z$) parts of the evolution for $N_m$.
Inserting (\ref{Nmrewrite}) in (\ref{BE}) we obtain the set of equations
necessary to calculate the response in linear approximation: 
\begin{eqnarray}
\label{eq14Gar}
\nonumber
P_{m+1|m} - P_{m-1|m}
&=&
\iu \w q_m
+
P_{m+1|m}( q_{m+1}-q_{m})
\\
&+&
P_{m-1|m}( q_{m-1}-q_{m})
\end{eqnarray}
%
here $\Pmmprima \equiv \Pmmprima (B_z = B_z^{0})$. 
%
%
To write (\ref{eq14Gar}) in terms of  $P_{m+1|m}$ and
$P_{m-1|m}$ the detailed balance equation
(\ref{detailed})  and the equality 
$\e^{\tf_{m'\,m}}N_{m'}^{(0)} = N_{m}^{(0)}$
have been utilized.

We are interested in the low frequency
behavior, so we expand
\begin{equation}
\label{expanqm}
q_m(\w)
\cong
q_m^{(0)}
+
\w
q_m^{(1)}
.
\end{equation}
We substitute (\ref{expanqm}) in (\ref{eq14Gar}), and we solve it
perturbatively in $\w$.
The
zero-frequency order gives:
\begin{eqnarray}
\label{eq14Garzero}
\nonumber
P_{m+1|m} - P_{m-1|m}
&=&
P_{m+1|m}( q_{m+1}^{(0)}-q_{m}^{(0)})
\\
&+&
P_{m-1|m}( q_{m-1}^{(0)}-q_{m}^{(0)})
.
\end{eqnarray}
Since (\ref{eq14Garzero}) depends only on the differences
$q_{m+1}^{(0)}-q_{m}^{(0)}$, and defining
\begin{equation}
\label{barqmdef}
\bar q_m
\equiv
q_{m+1}^{(0)} - q_{m}^{(0)},
\end{equation}
equation (\ref{eq14Garzero}) can be casted in a two-term
 recurrence.  With the ``boundary'' condition $P_{-S-1|-S} = 0$ the
 solution reads:
\begin{equation}
\label{barqm}
\bar q_m = 1
\end{equation}
and accordingly to (\ref{barqmdef}) and~(\ref{barqm}),
\begin{equation}
\label{qmzero}
q_m^{(0)}
=
m
-
\Mz
\end{equation}
where $\Mz$ is the magnetization. To obtain
(\ref{qmzero}) the normalization condition:
\begin{eqnarray}
\label{sumqm0}
\sum_{m=-S}^{S} N_m^{(0)} q_m^{(0)} &=& 0,
\end{eqnarray}
%
%
has been used. The above condition comes from $\sum_m N_m^{(1)}=0$.
Besides this follows from the unicity of the Fourier expansion of the norm and 
$
 \sum_m N_m^{(0)}=1
$
plus
$
\sum_m N_m = 1 
$
.
Now, we write
(\ref{eq14Gar}) up to first order in $\w$,
\begin{equation}
\label{eq14Garfirst}
0
=
\iu \w
q_m^{(0)}
+
P_{m+1|m}( q_{m+1}^{(1)}-q_{m}^{(1)})
+
P_{m-1|m}( q_{m-1}^{(1)}-q_{m}^{(1)})
.
\end{equation}
%
%
Once more, (\ref{eq14Garzero}) depends only
on the differences.  Introducing,
\begin{equation}
\label{rmdef}
r_m
\equiv
P_{m+1|m} e^{-\beta\varepsilon_m}
(q_{m+1}^{(1)}-q_{m}^{(1)})
, 
\end{equation}
%
%
equation~(\ref{eq14Garfirst}) transforms onto a two term recurrence relation,
obtaining for the $r_m$ elements,
%
%
\begin{equation}
\label{rm}
r_m
=
-\iu
\w
\sum_{j=-S}^{m}
q_j^{(0)} e^{- \beta \varepsilon_j}
\end{equation}
where we have again invoked $P_{-S-1|-S}=0$.  Now, (\ref{rm}) with
(\ref{rmdef}) and the first-$\w$ order normalization condition: 
\begin{eqnarray}
\label{sumqm1}
\sum_{m=-S}^{S} N_m^{(0)} q_m^{(1)} &=& 0,
\end{eqnarray}
%
permit to write $q_m^{(1)}$ as
%
\begin{equation}
\label{qm1}
\iu q_m^{(1)}
=
\sum_{i=-S}^{m-1}
\frac {\Phi_ i}{P_{i+1|i} N_i^{(0)}}
-
\sum_{j=-S}^{S}
N_j^{(0)}
\sum_{i=-s}^{j-1}
\frac {\Phi_ i}{P_{i+1|i} N_i^{(0)}}
\end{equation}
%
%
%
where $\Phi_i$ has been defined in (\ref{phim}).
The low-$\w$
expansion~(\ref{suslowomega}) and
(\ref{Nmrewrite}) allows to express $\tint$ as,
\begin{equation}
\label{iotint}
\iu
\tint
=
\frac {\beta}{
\suseq}
\sum_{m=-S}^{S}
m
N_m^{(0)} q_m^{(1)}.
\end{equation}
%
Introducing (\ref{qm1}) in
(\ref{iotint}) and using the equalities,
\begin{equation}
\label{equatity1tint}
\sum_{j=-S}^{S}
N_j^{(0)}
\sum_{i=-s}^{j-1}
\frac {\Phi_ i}{P_{i+1|i} N_i^{(0)}}
=
\sum_{i=-S}^{S-1}
\frac {\Phi_ i}{P_{i+1|i} N_i^{(0)}}
\sum_{j=i+1}^{S}
N_j^{(0)}
\end{equation}
and,
\begin{equation}
\label{equality2tint}
-\Mz
\sum_{j=i+1}^{S}
N_j^{(0)}
+
\sum_{j=i+1}^{S}
j N_j^{(0)}
=
\Phi_i
\end{equation}
we finally obtain the formula~(\ref{tintformula}) for the integral   
relaxation time.


{\it (ii) $\tintN$}

The time constant $\tintN$ arises when instead of the magnetization
susceptibility, one considers the low-$\Omega$ response of the
difference of the well populations $\Delta$
%
\begin{equation}
\label{delta}
\Delta 
\equiv
N_{+}
-
N_{-}
=
\sum_{m=-S}^{S}
\sgn (m - \mb)
N_m
,
\end{equation}
%
with $\varepsilon_{m_o}$ the maximum level.
Notice that the derivation of $\tintN$ must follow the same steps that for the
derivation of $\tint$.
Further taking into account the definition of $\Delta$ and from the
expression for $\susD$ in (\ref{susDlowomega}) $\tintN$ is given in
terms of $q_m^{(1)}$ (\ref{qm1}) as:
%
\begin{equation}
\label{tintNqm1}
\iu
\tintN
=
\frac {1}{\chi_{\Delta}}
\sum_{m=-S}^{S}
\sgn (m -\mb)
N_m^{(0)}
q_m^{(1)}
\end{equation}
%
Finally, using equalities~(\ref{equatity1tint}) and 
%
\begin{equation}
\label{equatity2tintN}
-\Delta
\sum_{j=i+1}^{S}
N_j^{(0)}
+
\sum_{j=i+1}^{S}
\sgn (m - \mb)
N_j^{(0)}
=
\Phi_j^{N}
,
\end{equation}
%
where $\Phi_j
^{N}$ is defined in (\ref{phimN}), formula~(\ref{tintN}) is readily obtained.


{\it (iii) $\tef$}

First we notice that the definition
for $\tef$ (\ref{sushighomega}) can be rewritten in terms of $N_m(t)$ as:
\begin{eqnarray}
\label{tefprevious}
\tef^{-1}
=
\frac 
{1}
{\Delta \Mz (0)}
\sum_{m=-S}^{S} m \dot N_m(0)
.
\end{eqnarray}
%
$ \dot N_m (0) $, is obtained directly
from the balance equation making $t = 0$ in (\ref{BE}).
Now, we recall that the system is initially at equilibrium with field $B_z =
B_z^{0} + \delta B_z $.
%
%
Up to first order in $\delta B_z$ this can be written:
\begin{equation}
\label{Nm(0)}
N_m (0)
=
N_m^{(0)}
+
\delta \Bz
\beta 
N_m^{(0)}
(m - \Mz).
\end{equation}
%
Next, we insert (\ref{Nm(0)}) in (\ref{BE}).
Considering: 
%
\begin{equation}
P_{m \vl m \pm 1}
N_{m \pm 1}^{(0)}
=
P_{m \pm 1 \vl m}
N_{m}^{(0)},
\end{equation}
%
(\ref{tefprevious}) reads:
%
\begin{eqnarray}
\label{tefendstep}
\nonumber
\tef^{-1}
=
\frac {\beta}{\suseq}
&\Bigg (&
\sum_{m = -S}^{S}
m
P_{m \vl m-1}
N_{m-1}^{(0)}
\\
&-&
\sum_{m = -S}^{S}
m
P_{m+1 \vl m}
N_{m}^{(0)}
\Bigg )
\end{eqnarray}
%
Finally, performing the change $m \to m +1$ in the first summand we obtain
the final formula~(\ref{tef}) for the inverse of the effective time.

\section{APPROXIMATE FORMULAS IN THE HIGH BARRIER CASE}
\label{app:AFHBC}

Here we derive approximate expressions for $\Deltauno$, $\eival_1$ and
$\eivalW$ in the high barrier case.

\begin{figure}[!tbh]
\centerline{
\resizebox{9.cm}{!}{%
\includegraphics[angle=-90]{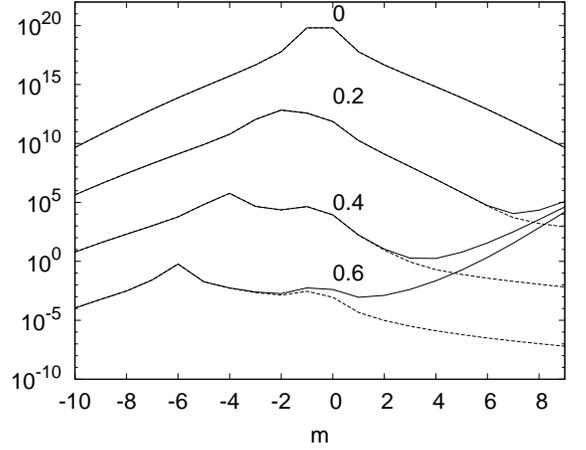}
}
}
\caption{Summands of $\tint$ (\ref{tintformula}), 
$\frac{\Phi^2_m}{N_m^{(0)}P_{m+1|m}}$,
 (solid line) and
   $\tintN$ (\ref{tintN}),
 $S \frac{\phim \phimN}{N_m^{(0)}P_{m+1|m}}$
 (dashed line).
Different $\hef$ are drawn: $\hef = 0. 0.2, 0.4, 0.6$.  
The rest of parameters are $S = 10$, $\sigma = 15$, $\lambda = 10^{-9}$. 
}
\label{fig:integrand}
\end{figure}

The technical dificulties deal mainly with the calculation of the sums in
$\tint$, $\tintN$ and $\tef$.
For that let us first consider the main approximation to handle with $\phim
^2$ and $\phim \phimN$.
In the range considered $\phim$ can be well approximated by:
%
\begin{eqnarray}
\label{phim_approx}
\nonumber
\phim
&\cong& 
\Phi_B
;
\quad
m < S -2
\\
\phim
&\cong &
\Phi_B + \Phi_W
;
\quad
m \cong S-1
\end{eqnarray}
where 
$\Phi_B \cong \sum_{j=-S}^{\mb} ( \Mz - j) N_j^{(0)} = \Phi_{\mb}$ 
and 
$\Phi_W \cong \sum_{j= \mb+1}^{S-1}( \Mz - j) N_j^{(0)} = \Phi_{S-1} -
\Phi_{\mb}$.
Besides there are two main contributions in the sum for $\tint$, one around
$\mb$ and the other one around $m \cong S-1$ (see Fig.~\ref{fig:integrand}).
This last peak in the sum is due to $\Phi_W$.
On the other hand the quantity $\phim \phimN$ results well approximated for
practical purposes by the constant:
%
\begin{equation}
\label{phimN_approx}
\phim
\phimN
\cong
\Phi_B
\Phi ^N
\end{equation}
%
with 
$\Phi ^N = \sum_{j=-S}^{\mb} (\Delta + 1) N_j^{(0)} = \Phi^N_{\mb}$.
Then the sum for $\tintN$ has only one main
contribution around $\mb$ (see Fig.~\ref{fig:integrand}).


Now, we give the main steps to calculate the formula
(\ref{Deltaunohighsigma}) for $\Deltauno$.
In the bimodal approximation $\tint \cong
\Deltauno/\eival_1 + (1 - \Deltauno)/\eivalW$.
Then, and taking into account that $\eivalW \gg \eival_1$ and $\eival_1
\cong \tintN$:
%
\begin{equation}
\label{Deltaunoapproxprev}
\Deltauno
\cong
\frac {\tint}{\tintN}
.
\end{equation}
%
Next we use approximations~(\ref{phim_approx}) and~(\ref{phimN_approx}).
Neglecting  the second peak in the sum, which is produced by  intrawell dynamics (it would contribute to
the summand $(1 - \Deltauno) / \eivalW$) we write:
%
\begin{equation}
\label{DeltaunophiphiN}
\Deltauno
\cong 
S
\frac {\chi_\Delta}{\suseq}
,
\end{equation}
%
here we have also used that in the high barrier limit $\Phi ^ B = S
\Phi ^N$ (see Fig.~\ref{fig:integrand}).
Notice that (\ref{DeltaunophiphiN}) depends only on equilibrium magnitudes.
To compute $\suseq$ and $\chi_{\Delta}$ we approximate the partition function
by:
%
\begin{equation}
\label{Z_approx}
\mathcal Z
\cong 
\e^{\sigma}
\left \{
2 \cosh \xi
+
\e^{\sigma ( \factorq ^2 -1 )}
\cosh \factorq \xi
\right \},
\end{equation}
with $\factorq = (S-1)/S$.
This form for $\mathcal Z$ has been calculated considering only the states
$m = \pm S$ and $m = \pm (S-1)$.
Restricting ourselves to the lowest states, becomes justified when $\sigma \gg
1$.
However, at sufficiently high $\hef$, levels in the right
well with $m < S -1$ have less energy than $m = -S +1$ even $m = -S$, then
they would be more populated.
On the other hand,  this is
the minimal model which includes sufficient states to deal with intrawell
and overbarrier contributions.
Then, from (\ref{Z_approx}) we obtain $\suseq$ and $\chi_{\Delta}$.
Finally, considering only the leading terms in $\sigma$ we obtain the final
expression (\ref {Deltaunohighsigma}) for $\Deltauno$.


Now we turn our attention to $\eival_1$.
In the main text we have argued that $\eival_1 \cong \tintN^{-1}$.
Then, and considering (\ref{phimN_approx}), we write:
%
\begin{equation}
\label{eival_1_first}
\eival_1^{-1}
\cong
\tintN
\cong 
\frac {\beta}{\chi_{\Delta}}
\Phi \Phi^N
\sum_{m=-S}^{S-1}
\frac {1}
{N_m^{(0)}P_{m+1|m}}
.
\end{equation}
We use now $\theta_m$ [see Eq.~(\ref{wurger})],
{\it i.e}
%
$
\theta_m 
=
\sum_{j=m+1}^{S}
N_j^{(0)}
$
;
%
then it can be checked that:
\begin{equation}
 \Delta \cong  2 \theta_{\mb} -1
\end{equation}
and
\begin{equation}
\chi_{\Delta}  \cong 2 \partial_{\Bz} \theta_{\mb}
= 2 \beta \Phi^B
\end{equation}
%
Thus $\eival_1^{-1}$ reads:
%
\begin{equation}
\label{eival_1approxprev}
\eival_1^{-1}
\cong
( 1 - \theta_{\mb} ) \theta_{\mb}
\sum_{m=-S}^{S-1}
\frac {1}
{N_m^{(0)}P_{m+1|m}}
.
\end{equation}
%
First,  Eq.~(\ref{eival_1approxprev}) equals the
W\"urger result~(\ref{wurger}),
since $\theta_m \cong \theta_{\mb} = 1/2$.
Furthermore, the sums for $\theta_{\mb}$ and $\sum 1 / N_m^{(0)} P_{m+1 \vl
  m}$ can be evaluated following Garanin in some limiting cases \cite{gar97}.
In particular the classical limit can be carried out.
In this limit the sum is  converted into an integral, which in turn
can be solved.
Then using approximate expressions for the classical $\mathcal Z$ when
$\sigma \gg 1$ (see Sec.~II of \cite{gar2000}) and the same for
$\theta_{\mb}$, we recover the Brown result~(\ref{Brown_result}) for
$D(z)\propto(1-z^2)$ [see Eq.~(\ref{f-pclassical})].
Eventually, one also considers $\sigma /S > 1$ to evaluate the sums in
(\ref{eival_1approxprev}).
In this case, becomes sufficient consider only the two main terms in the sum,
namely  
$1 / N_m^{(0)} P_{\mb+1 \vl \mb}$
and
$1 / N_m^{(0)} P_{\mb-1 \vl \mb}$.
In addition we only take into account the leading terms in $\sigma$ for
$\theta_{\mb}$, obtaining the Arrhenius formula~(\ref{Arrhenius}) for $\eival
_1$.


Finally we consider $\eivalW$.
In the expresion for $\eivalW$ [Eq.(\ref{LambdaWiden})], the
denominator can be approximated by $\eival_1 \tint -1 \cong \Deltauno
-1$ [see Eq.~(\ref{Deltaunoapproxprev})].  Finally assuming $\tef
^{-1} \gg \eival _1$ (see Fig.~\ref{fig:taus}) $\eivalW$ reads:
%
\begin{equation}
\label{LambdaWapproxprev}
\eivalW
\cong 
\frac {\tef^{-1}}{1 - \Deltauno}
\end{equation}
%
Then $\eivalW / \eival_1$:
%
%
\begin{eqnarray}
\nonumber
\frac {\eivalW}{\eival_1}
=
\frac {1}{4 S^2}
\frac {\Deltauno}{1 - \Deltauno}
&\Bigg (&
\sum_{m=-S}^{S-1}
N_m^{(0)}P_{m+1|m}
\Bigg )
\\
&\times&
\Bigg (
\sum_{m=-S}^{S-1}
\frac {1}
{N_m^{(0)}P_{m+1|m}}
\Bigg ).
\end{eqnarray}
To obtain this formula we have used equation~(\ref{eival_1_first}) for $\eival
_1$ and~(\ref{DeltaunophiphiN}) for $\Deltauno$ together with $\Phi_B \cong
S \Phi ^N$.
Now, taking only leading terms in the sum we readily obtain
(\ref{eival_1/eival_W}) in the regime of interest.


\section {EXACT ANALYTICAL SUSCEPTIBILITY FOR S=1}
\label{app:S1}

When $S=1$ it is possible to derive an exact analytical expression for $\sus
$.
%
%
%
For that, we calculate the eigenvalues of the, in this case, $3 \times 3$
relaxation matrix $\mathcal R$, obtaining:
%
\begin{equation}
\label{modesS1}
\eival_{1,2}
=
(\Gamma_{+}
+
\Gamma_{-}
)
\mp
\sqrt
{
(\Gamma_{+}
-
\Gamma_{-}
)^2
+
4
P_{1 \vl 0}^{(0)} P_{-1 \vl 0}^{(0)}
}
,
\end{equation}
plus the zero eigenvalue, $\eival_0 = 0$. 
Here $\Gamma_{\pm} \equiv P_{\pm 1 \vl 0}[1 + \exp \{ \beta
(\epsilon_{\pm 1} - \epsilon_0)\}]$
Notice that the bimodal description is exact for $S=1$ since $\eivalW \equiv
\eival_2$.
Besides, using the relation of $\tef^{-1}$ with the eigenvalues, {\it
i.e.}  $\tef^{-1} = \Deltauno \eival_1 + (1 - \Deltauno) \eival_2 $
[Eq.(\ref{tintauto:tefauto})] $\Deltauno$ yields:
%
\begin{equation}
\label{DeltaunoS1}
\Deltauno
=
\frac {1}{\eival_2 - \eival_1}
\left (
\eival_2 
- 
\tef^{-1}
\right )
,
\end{equation}
%
with $\tef^{-1} = \beta / ( \suseq \mathcal Z ) (P_{1 \vl 0}^{(0)} + P_{-1
  \vl 0}^{(0)})$.

The two modes  $\eival_1$, $\eival_2$ and $\Deltauno$
are introduced in  (\ref{susapprox}) obtaining the
exact expression for the response.


                       


\bibliography{/home/david/notas_articulos/cosas_tex/david}

\begin{thebibliography}{26}
\bibitem {fn:mb} \bibnamefont{The maximum level is given by $m_{\protect \mathrm {b}}= -B_z/2D + F$, with $F =\protect \mathcal F [ B_z/2D + S + 1/2 ] - 1/2$ and $\protect \mathcal F [x]$, the fractional part of $x$.}
\bibitem  {fn:hi} \bibnamefont {For $S$ half-integer, at $h=0$ the $(2S +1)$ states are double degenerated: $|\Lambda _0 \delimiter "526930B = |\Lambda _1 \delimiter "526930B = 0$ ($P_{-1/2 |1/2}=0$). The rest of eigenvalues describe symmetrical proccesses in both wells analogous to the ones explained for $S$ integer. }
\bibitem {fn:minima} \bibnamefont{ At $h= (2 n + 1) / (2S - 1)$ the lowest eigevalue goes to zero. At these bias $m = -n$ and $m = -n +1$ becomes degenerate, thus the relaxation time diverges, since $P_{-n |-n + 1}(0) = 0$ in the model used. }
\end{thebibliography}


\end{document}